\documentclass[prx,amsmath,amssymb, notitlepage,superscriptaddress,nofootinbib,longbibliography]{revtex4-2}

\pdfoutput=1

\usepackage{dsfont}
\usepackage{float}
\usepackage{array}
\usepackage{slashed,bbold}

\usepackage{amsmath,amssymb,bm} 
\usepackage{graphicx}

\usepackage[tight]{subfigure} 

\usepackage{color} 
\usepackage[papersize={8.5in,11in}]{geometry}

\usepackage{color}
\definecolor{darkblue}{rgb}{0.,0.,0.4}
\definecolor{darkred}{rgb}{0.5,0.,0.}
\definecolor{BlueViolet}{RGB}{138,43,226}
\definecolor{SkyBlue}{RGB}{30,144,255}
\definecolor{DarkGreen}{RGB}{0,100,0}
\usepackage[pdftex,colorlinks=true,linkcolor=darkblue,citecolor=blue,urlcolor=darkred]{hyperref}
\usepackage[normalem]{ulem}

\usepackage{amsmath,amssymb,bm} 
\usepackage{graphicx}

\usepackage[tight]{subfigure}

\def \nn{\nonumber \\}
\begin{document}

\title{Tunneling in Fermi Systems with Quadratic Band Crossing Points}

\author{Ipsita Mandal}

\affiliation{Faculty of Science and Technology, University of Stavanger, 4036 Stavanger, Norway}
\affiliation{Nordita, Roslagstullsbacken 23, SE-106 91 Stockholm, Sweden}

\begin{abstract}
We investigate the tunneling of quasiparticles through a rectangular potential barrier of finite height and width, in 2d and 3d semimetals with band structures consisting of a quadratic band crossing point. We compute the transmission coefficient at various incident angles, and also demonstrate the behaviour of the conductivity and the Fano factor.
We discuss the distinguishing signatures of these transport properties in comparison with other semimetals, as well as electrons in normal metals.
\end{abstract}
\maketitle

\tableofcontents
%==========================================================================

\section{Introduction}
%%%%%%%%%%%%%%%%%%%%%%%%%%%%%%%%%%%%%

%%%%%%%%%%%%%%%%%%%%%%%%%%%%%%%%%%%%%%%%%%%%%%%%%%%
\begin{figure}[htb]
{\includegraphics[width = 0.45 \textwidth]{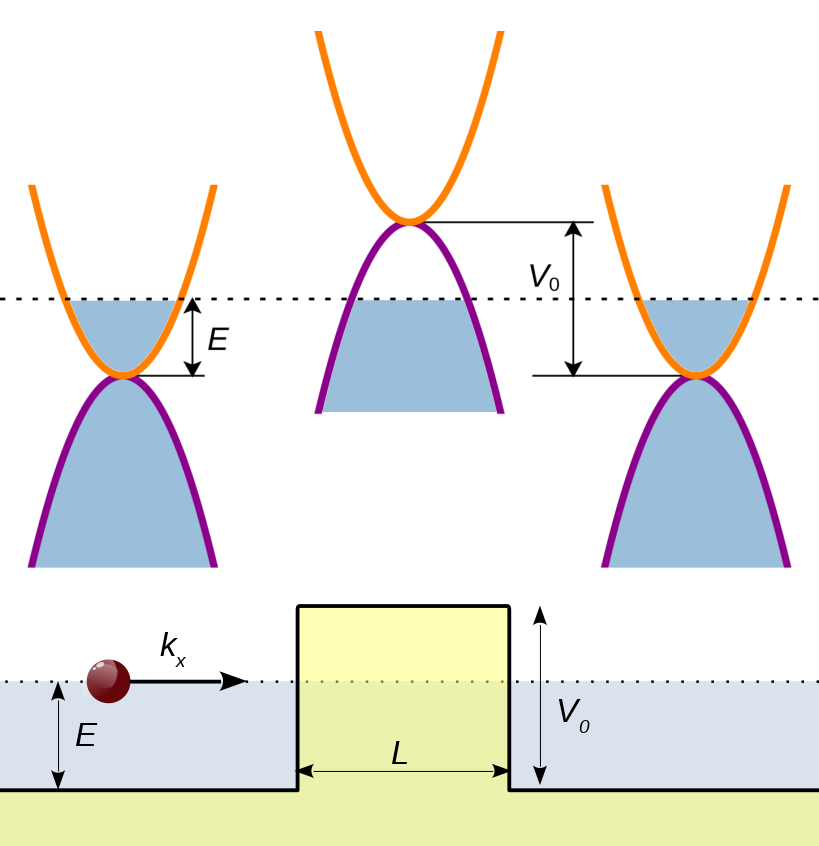}}
\caption{Tunneling through a potential barrier in a QBCP material. The upper panel shows the schematic diagrams of the spectrum of quasiparticles about a QBCP, with respect to a potential barrier in the $x$-direction. The lower panel represents the schematic diagram of the transport across the potential barrier. The Fermi level (indicated by dotted lines) lies in the conduction band outside the barrier, and in the valence band inside it. The blue fillings indicate occupied states.}
\label{figbands}
\end{figure}
%%%%%%%%%%%%%%%%%%%%%%%
% https://arxiv.org/pdf/cond-mat/0604323.pdf

Multiband fermionic systems may exhibit a band crossing point in the Brillouin zone where two or more bands cross. If the chemical potential is adjusted to lie exactly at that point, the Fermi surface shrinks to a Fermi node. The most famous example of such a Fermi node is the case of a linear band crossing, whose low energy properties are described by Dirac fermions, and are conspicuous in systems like nodal superconductors and graphene.
In this paper, we consider systems with a quadratic band crossing point (QBCP) somewhere in their two-dimensional (2d) \cite{kai-sun,tsai,ips-seb} or three-dimensional (3d) \cite{Abrikosov,LABIrridate,MoonXuKimBalents} Brillouin zones. 2d QBCPs can be realised in
checkerboard \cite{kai-sun} (at $1/2$ filling), Kagome \cite{kai-sun} (at $1/3$ filling), and Lieb \cite{tsai} lattices.
On the other hand, pyrochlore iridates $\text{A}_ 2\text{Ir}_ 2\text{O}_ 7$ (A is a lanthanide element~\cite{pyro1,pyro2}) have been shown to host a 3d QBCP. Such bandstructures have also been realised in 3d gapless semiconductors in the presence of a sufficiently strong spin-orbit coupling \cite{Beneslavski}, such that the resulting model of a semimetal is indeed relevant for materials like gray tin ($\alpha$-Sn) and mercury telluride (HgTe). These systems are also known as ``Luttinger semimetals"~\cite{igor16} due to the fact that the low-energy fermionic degrees of freedom are captured by the Luttinger Hamiltonian  of inverted band-gap semiconductors \cite{luttinger,murakami}.

Our aim is to compute the tunneling coefficients and other transport characteristics when the quasiparticles of the QBCP
semimetals are subjected to a potential barrier of finite strength and width along one direction, which is chosen to be the $x$-axis. This scenario is represented in the cartoon in Fig.~\ref{figbands}. Our results will show how these transport characteristics are significantly different from those in normal metals, due to the presence of multiple bands. We
will also compare their features with those of other semimetals like graphene, bilayer graphene and three-band pseudospin-$1$ systems.

The paper is organized as follows. In Sec.~\ref{sec2dmodel}, we study the 2d QBCPs, while Sec.~\ref{sec3dmodel} deals with the 3d QBCP case. We compare our findings with the results for some other known bandstructures in Sec.~\ref{seccompare}. Finally, we end with a summary and outlook in Sec.~\ref{secsum}.

%%%%%%%%%%%%%%%%%%%%%%%%%%%%%%%%%%%%%%
\section{2d Model}
\label{sec2dmodel}

For a 2d system, the particle-hole symmetric QBCP with $C_6$ rotational symmetry is described by the Hamiltonian \cite{kai-sun}:
\begin{align}
\mathcal{H}_{2d}^{kin}(k_x,k_y)&=
 \frac{\hbar^2 }{2\,m}
 \Big [ \, 2\, k_x\,k_y\, \sigma_x +  \left( k_y^2-k_x^2\right) \sigma_z  \,\Big ]
\end{align} 
in the momentum space,
with eigenvalues
\begin{align}
\varepsilon_{2d}^{\pm}(k_x,k_y) = \pm  \frac{\hbar^2 \left( k_x^2 + k_y^2 \right)}{2\,m} ,
\end{align}
where the ``$+$" and ``$-$" signs, as usual, refer to the conduction and valence bands respectively.
The corresponding eigenvectors are given by:
\begin{align}
\Psi_+ =\frac{1}{\sqrt{k_x^2+k_y^2}} \lbrace k_y, k_x\rbrace\,, \text{ and }
  \Psi_- =\frac{1}{\sqrt{k_x^2+k_y^2}} \lbrace  -k_x, k_y  \rbrace,
\end{align}
respectively.

The 2d system is modulated by a square electric potential barrier of
height $V_0 $ and width $L$, giving rise to an $x$-dependent potential energy function:
\begin{align}
V ( x ) = \begin{cases} V_0 &  \text{ for } 0 < x < L \\
0 & \text{ otherwise} \,.
\end{cases}
\label{eqpot}
\end{align}
Hence, we need to consider the total Hamiltonian:
\begin{align}
\mathcal{H}_{2d}^{tot} &=
 \mathcal{H}_{2d}^{kin}(-\mathrm{i}\,\partial_x, -\mathrm{i}\,\partial_y)+V(x)
\end{align} 
in position space. We choose the $x$-axis along the transport direction, and place the chemical potential at an energy $E >0$ in the region outside the potential barrier. The Fermi energy $E$ can in general be tuned by chemical doping
or a gate voltage.

\subsection{Formalism}

For a material of a sufficiently large transverse dimension $W$, the boundary conditions should be irrelevant for the bulk response, and we use this freedom to simplify the calculation. Here, on a physical wavefunciton $\Psi^{\mathrm{tot}}$ we impose periodic boundary conditions:
\begin{align}
 \Psi^{\mathrm{tot}}(x,W) =
 \Psi^{\mathrm{tot}}(x,0) \,.
\end{align}
The  transverse momentum $k_y$ is conserved, and it is quantized due to the periodicity in the transverse width $W$, and hence takes the form:
\begin{align}
\quad k_y =\frac {2\,\pi\,n} {W} \equiv q_{n}\,,
\end{align}
where $n \in \mathbb{Z}$.
For the longitudinal direction, we seek plane wave solutions of the form $ e^{\mathrm{i}\,k_x x} $. Then the full wavefunction is given by:
\begin{align}
 \Psi^{\mathrm{tot}}(x,y,n) =
\text{const.}
\times \Psi_{ n}(x)\,  e^{ \mathrm{i}\,q_{n} y }\,,
\end{align}
For any mode of given transverse momentum component $k_y$, we can determine the $x$-component of the wavevectors of the incoming, reflected, and transmitted waves (denoted by $k_{\ell}$), by solving $
\varepsilon_{2d}^\pm(k_x ,  n ) =  \pm \frac{ \hbar^2 \left( k_{\ell}^2 + q_{n}^2  \right)} {2\,m}\,.$
In the regions $x<0$ and $x>L$, we have only propagating modes ($  k_\ell $ is real), while the $x$-components
in the scattering region (denoted by $ \tilde k  $),  are given by $\tilde k^2 = \frac{2\,m\,|E-V_0|} {\hbar^2} -q_n^2 $,
and may be propagating (imaginary part of $\tilde k $ is zero) or evanescent (imaginary part of $\tilde k $ is nonzero).

We will follow the procedure outlined in Refs.~\onlinecite{salehi,beenakker} to compute the transport coefficients. We consider the transport of positive energy states
($\Psi_+$) corresponding to electron-like particles. The transport of hole-like excitations ($\Psi_-$) will be similar. Hence, the Fermi level outside the potential barrier is adjusted to a value $E =\varepsilon_{2d}^+(k_x,k_y) $.
Such a scattering state $\Psi_{ n,+}$, in the mode labeled by $n$, is
constructed from the states:
\begin{align}
 \Psi_{ n} (x)=&  \begin{cases} \phi_L & \text{ for } x<0 \,, \\
 \phi_M & \text{ for } 0< x < L \,,\\
\phi_R &  \text{ for } x > L \,,
\end{cases} \nonumber \\
%%%%%%%%%%%%%%%%%%%%%%%%%%%%%%%%%%%%%%%%
 \phi_L = & \, \frac{ 
 \Psi_+ (  k_\ell,  q_{n}) \,  e^{\mathrm{i}\, k_\ell x }
+  r_{  n}\, \Psi_+ (  -k_\ell,  q_{n}) \,   e^{-\mathrm{i}\, k_\ell x }
} 
{\sqrt{ \mathcal{V} (k_\ell,  n)}}\, ,\nonumber \\
%%%%%%%%%%%%%%%%%%%%%%%%%%%%%%%%%%%%%%%%%%%%5
  \phi _M = &\,\Big[  \alpha_{ n}\,\Psi_+ (  \tilde k,  q_{n}) \,
 e^{\mathrm{i}\,\tilde k\,  x } 
 + \beta_{ n} \,\Psi_+ (  -\tilde k,  q_{n}) \,
 e^{-\mathrm{i}\,\tilde k  \, x } \Big]   \,\Theta\left( E-V_0  \right)
%  \nonumber\\
%& 
 + \Big[ \alpha_{  n}\,\Psi_- (  \tilde k,  q_{n}) \,
 e^{\mathrm{i}\,\tilde k\,  x } 
 + \beta_{n} \,\Psi_- ( -\tilde k,  q_{n}) \,
 e^{-\mathrm{i}\,\tilde k  \, x }   \Big]  \,\Theta\left( V_0-E  \right) ,\nonumber \\
 %%%%%%%%%%%%%%%%%%%%%%%%%%%%%%%%%%%%%%%%%%%%%%%%%%%%%%%%%%%55
\phi_R = & \,  t_{ n}\, \Psi_+ ( k_\ell,  q_{n})\,
 \frac{  e^{\mathrm{i}\, k_\ell \left( x-L\right)}} 
{\sqrt{ \mathcal{V} (k_\ell,  n)}}\,,\nonumber \\
%%%%%%%%%%%%%%%%%%%%%%%%%%%%%%
 \mathcal{V} (k_\ell,  n) \equiv &  \,  |\partial_{k_\ell} \varepsilon_+ (k_\ell, n)|
=  \frac{ \hbar^2  k_\ell } {m}\,,
\quad k_{\ell} = \sqrt{\frac{2\,m\,E} {\hbar^2}-q_n^2}
\,,\quad  \tilde k = \sqrt{\frac{2\,m\,|E-V_0|} {\hbar^2} -q_n^2}\,,
\end{align}
where we have used the velocity $ \mathcal{V} (k_\ell,  n)$ to normalize the incident, reflected and transmitted plane waves. 
Note that for $V_0 > E$, the Fermi level within the potential barrier lies within the valence band, and we must use the valence band wavefunctions in that region.

The boundary conditions can be obtained by integrating the equation $\mathcal{H}_{2d}^{tot}\, \Psi^{\mathrm{tot}} = E \, \Psi^{\mathrm{tot}} $ over a small interval in the $x$-direction around the points $x=0$ and $x=L$. The results are that the two components of the wavefunction be continuous at the boundaries. These conditions are sufficient to guarantee the continuity of the current flux along the $x$-direction
\footnote{From wavefunction matching, we have two equations from the two boundaries.
For 2d QPCB, each of these equations has two components as each wavevector has two components.
Therefore we have four equations for four undetermined coefficients.
We do not need to match the first derivatives of the wavefunction as those will be redundant equations.}.
In particular, the reflection and transmission amplitudes $r_n, t_n $, and the two coefficients
$\left( \alpha_n, \beta_n \right) $, are determined from these boundary conditions.
This mode-matching procedure gives us:
\begin{align}
r_n(E, V_0) & = \begin{cases}
\frac{\left(\tilde{k}^2 k_\ell^2- q_n^4 \right) 
\sin (\tilde{k} L)}{\left(\tilde{k}^2 k_\ell^2+ q_n^4\right) \sin (\tilde{k} L)
-2 \,\mathrm{i} \, \tilde{k}\, k_\ell \,q_n^2 \cos (\tilde{k} L)}
&\text{ for } E < V_0 \\
%%%%%%%%%%%%
\frac{ \left(k_{\ell}^2- \tilde{k}^2\right) \sin \left( \tilde{k} L \right ) }
{\left( \tilde{k}^2+ \, k_\ell^2\right) \sin \left(\tilde{k} L \right )
+2  \,\mathrm{i} \, \tilde{k}\, \, k_\ell \cos \left (\tilde{k} L \right)}
&\text{ for } E > V_0\,. 
\end{cases}
\label{eqrval}
\end{align}
and
\begin{align}
t_n(E, V_0) & = \begin{cases}
-\frac{2 \,\mathrm{i} \,\tilde{k}\, k_\ell \,q_n^2}
{\left(\tilde{k}^2 \,k_\ell^2+ q_n^4\right) 
\sin \left(\tilde{k} L \right )
-2 \,\mathrm{i}\, \tilde{k}\, k_\ell\, q_n^2 \cos \left (\tilde{k} L \right)}
&\text{ for } E < V_0 \\
%%%%%%%%%%%%
\frac{2 \,\mathrm{i} \,  \tilde k\, k_\ell }
{\left( \tilde{k}^2+ \, k_\ell^2\right) \sin \left(\tilde{k} L \right )
+2  \,\mathrm{i} \, \tilde{k}\, \, k_\ell \cos \left (\tilde{k} L \right)}
&\text{ for } E > V_0\,. 
\end{cases}
\label{eqtval}
\end{align}
The reflection and transmission coefficients at an energy $E$ are given by 
\begin{align}
R( E ,  V_0,\phi) = | r_n( E, V_0 )|^2 \text{ and }
T( E ,  V_0,\phi) = | t_n( E, V_0 )|^2 \,,
\end{align}
respectively, where $\phi = \tan^{-1} \left( \frac {q_{n}} {k_\ell} \right)$
is the incident angle of the incoming wave.

%%%%%%%%%%%%%%%%%%%%%%%%%%%%%%%%%%%%%%%%%%%%%%%%%%
\begin{figure}[htb]
\subfigure[]{\includegraphics[width = 0.14 \textwidth]{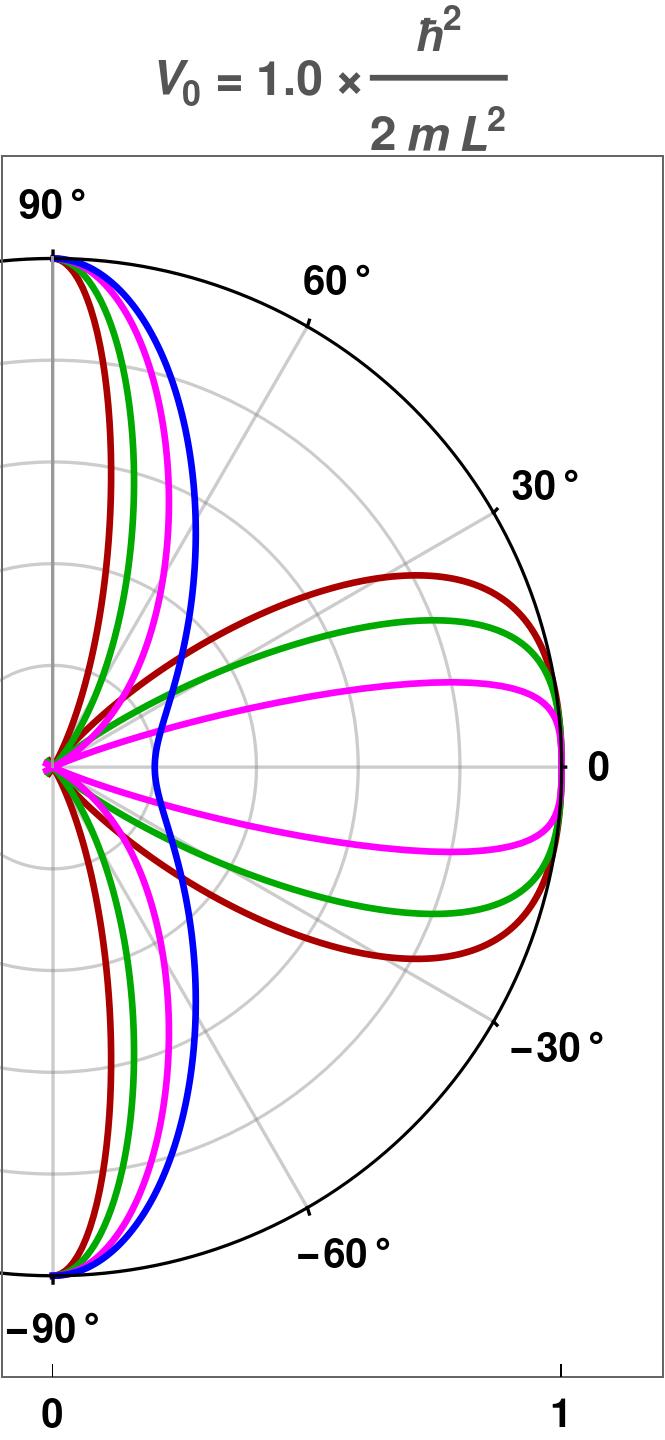}} \hspace{2 cm}
\subfigure[]{\includegraphics[width = 0.14 \textwidth]{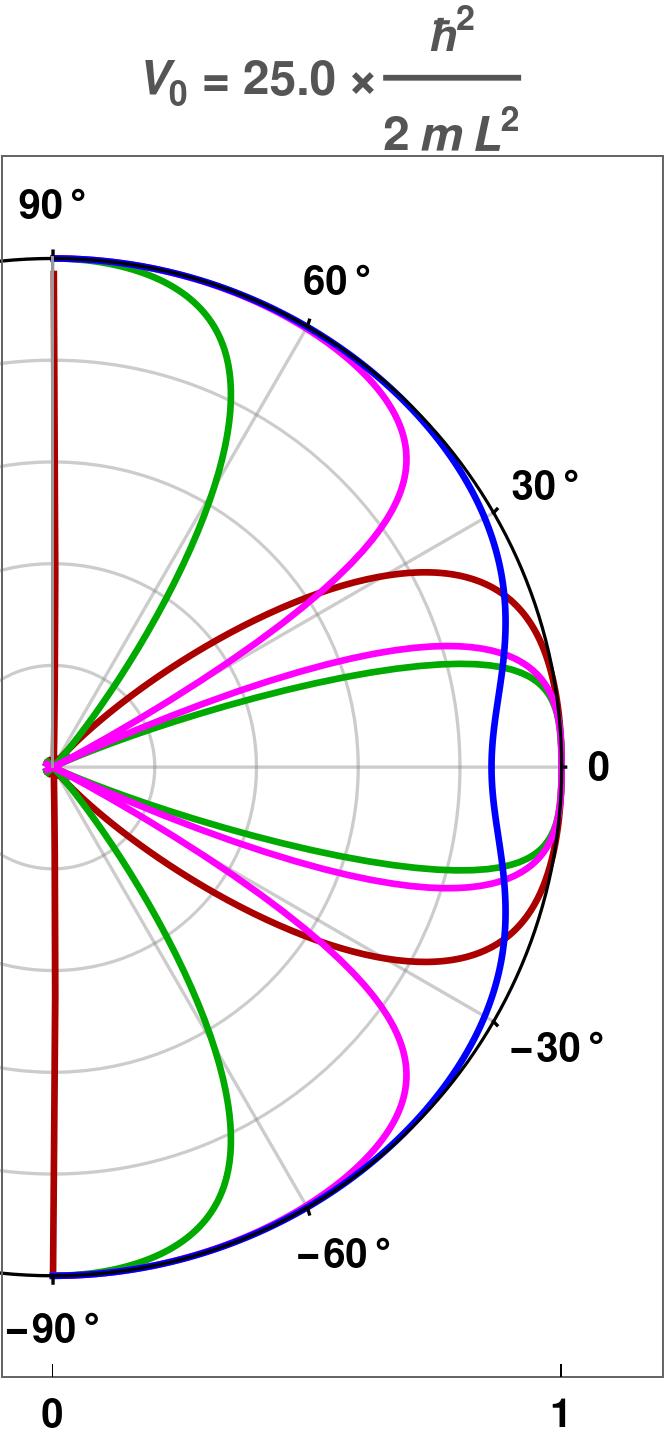}}\hspace{2 cm}
\subfigure[]{\includegraphics[width = 0.14 \textwidth]{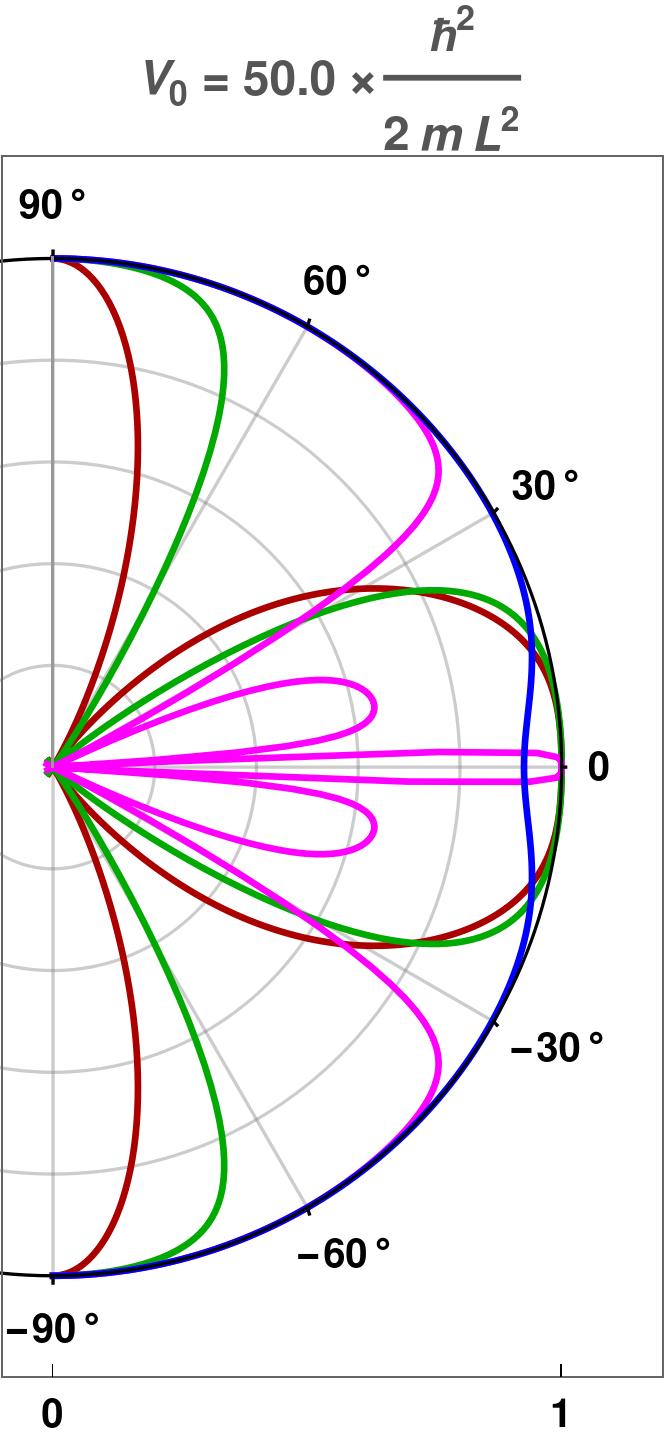}} \\
\subfigure[]{\includegraphics[width = 0.14 \textwidth]{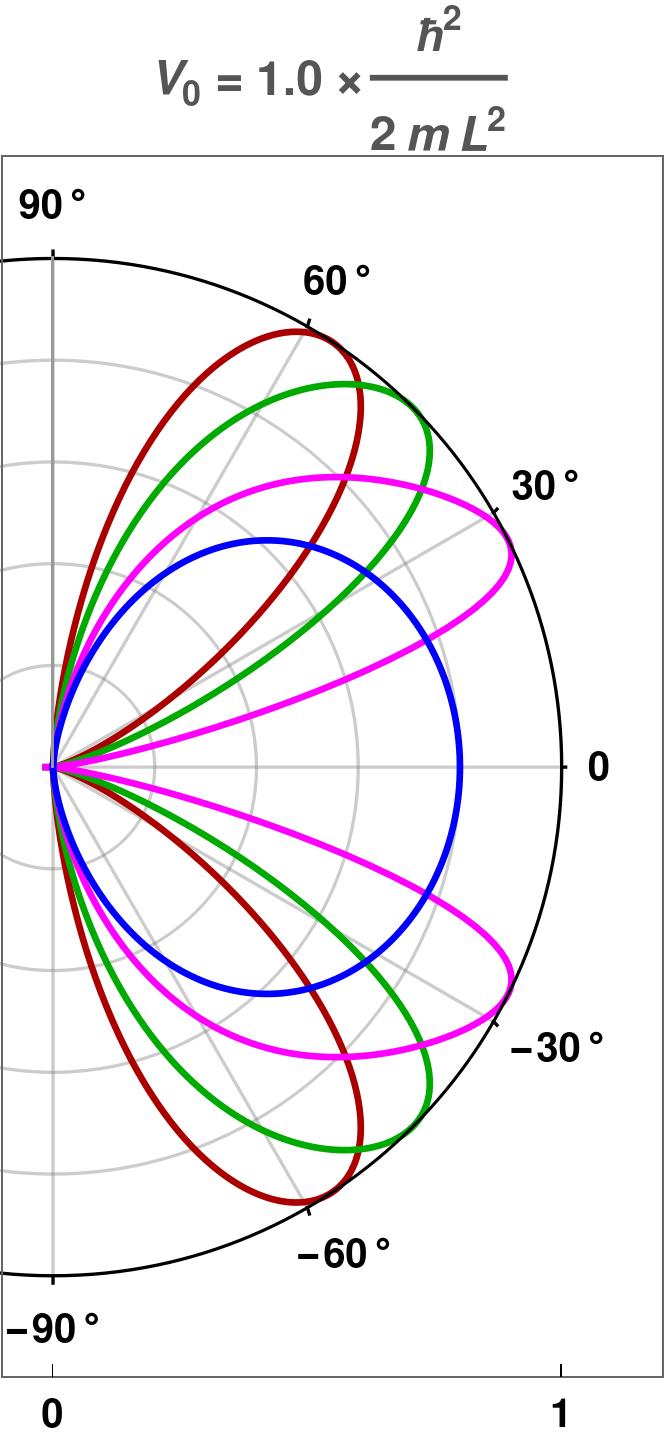}} \hspace{2 cm}
\subfigure[]{\includegraphics[width = 0.14 \textwidth]{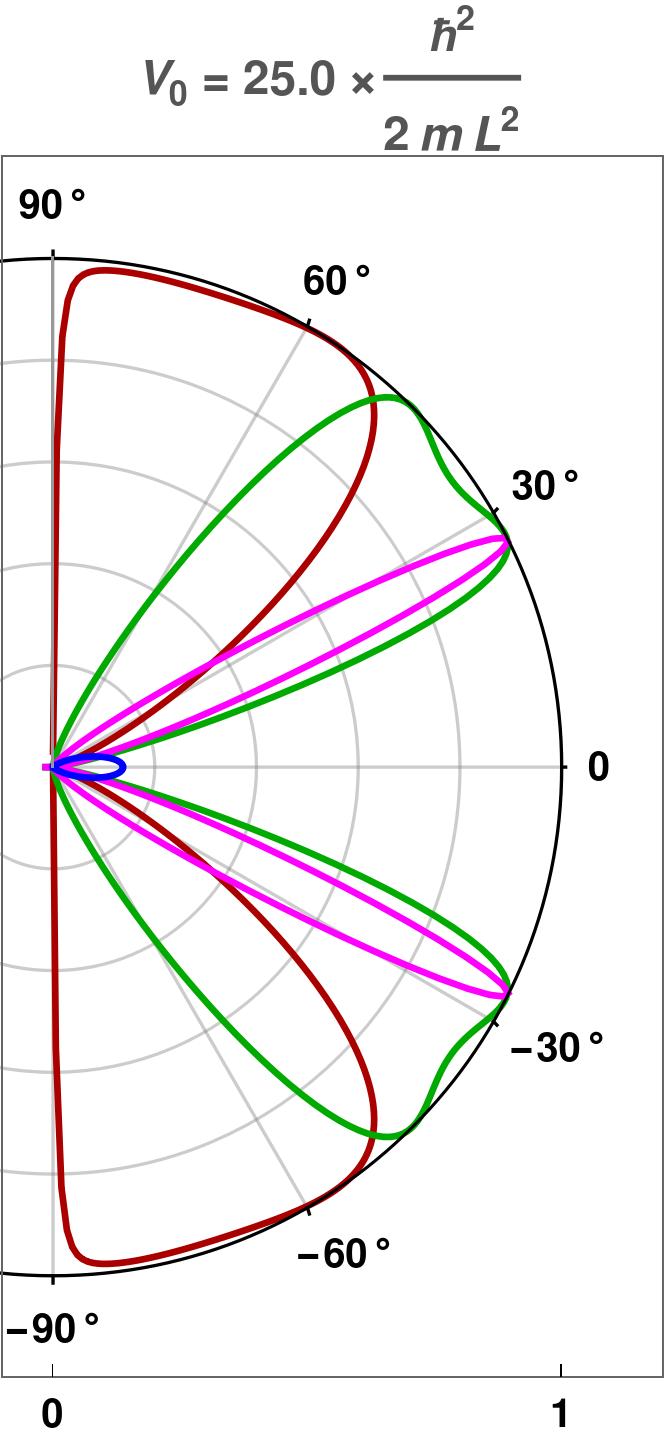}}\hspace{2 cm}
\subfigure[]{\includegraphics[width = 0.14 \textwidth]{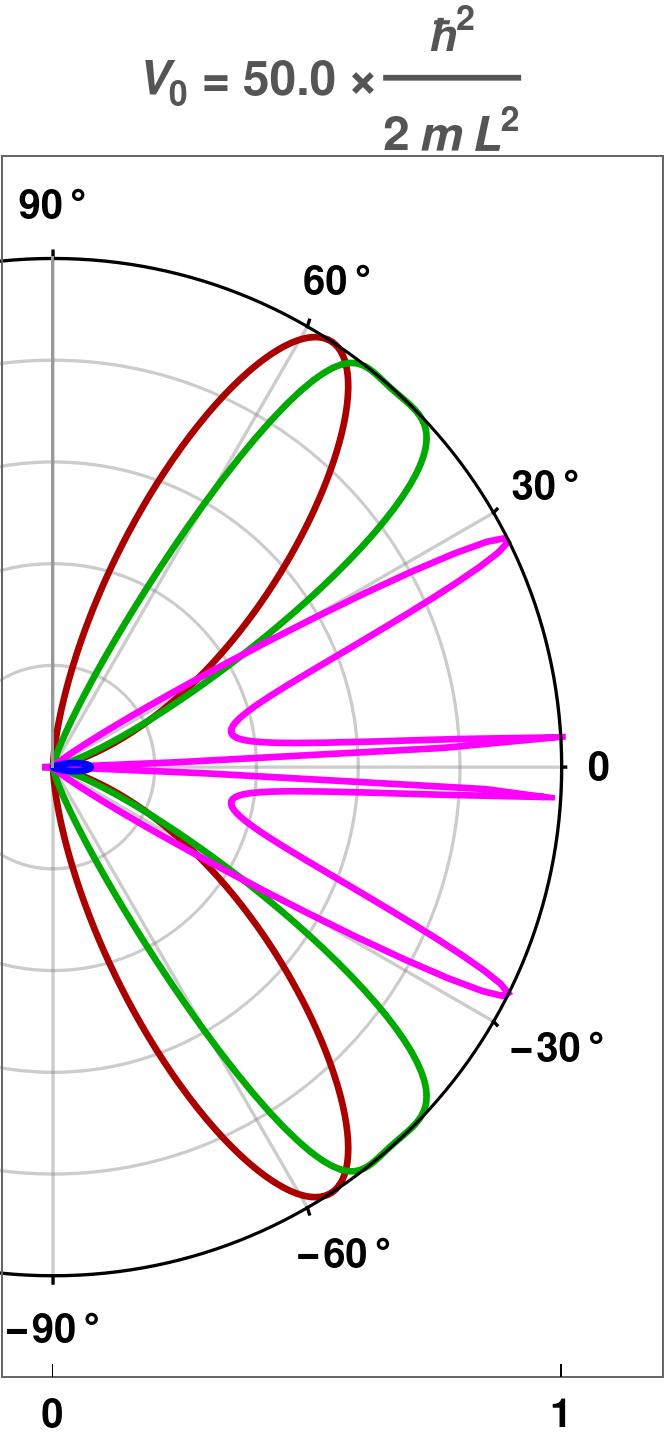}} 
\caption{2d QBCP: The polar plots show the reflection coefficient $ R(E, V_0,\phi) \big \vert_{E \leq V_0}$ and the transmission coefficient $T(E, V_0,\phi) \big \vert_{E \leq V_0}$ as functions of the incident angle $\phi$ for the parameters $E= 0.3 \,V_0$ (red), $E=0.5\,V_0$ (green), $E=0.8\,V_0$ (magenta)
and $E=1.0 \,V_0$ (blue).}
\label{figRTenLessV}
\end{figure}
%%%%%%%%%%%%%%%%%%%%%%%%

%%%%%%%%%%%%%%%%%%%%%%%%%%%%%%%%%%%%%%%%%%%%%%%%%%%
\begin{figure}[htb]\subfigure[]
{\includegraphics[width = 0.14 \textwidth]{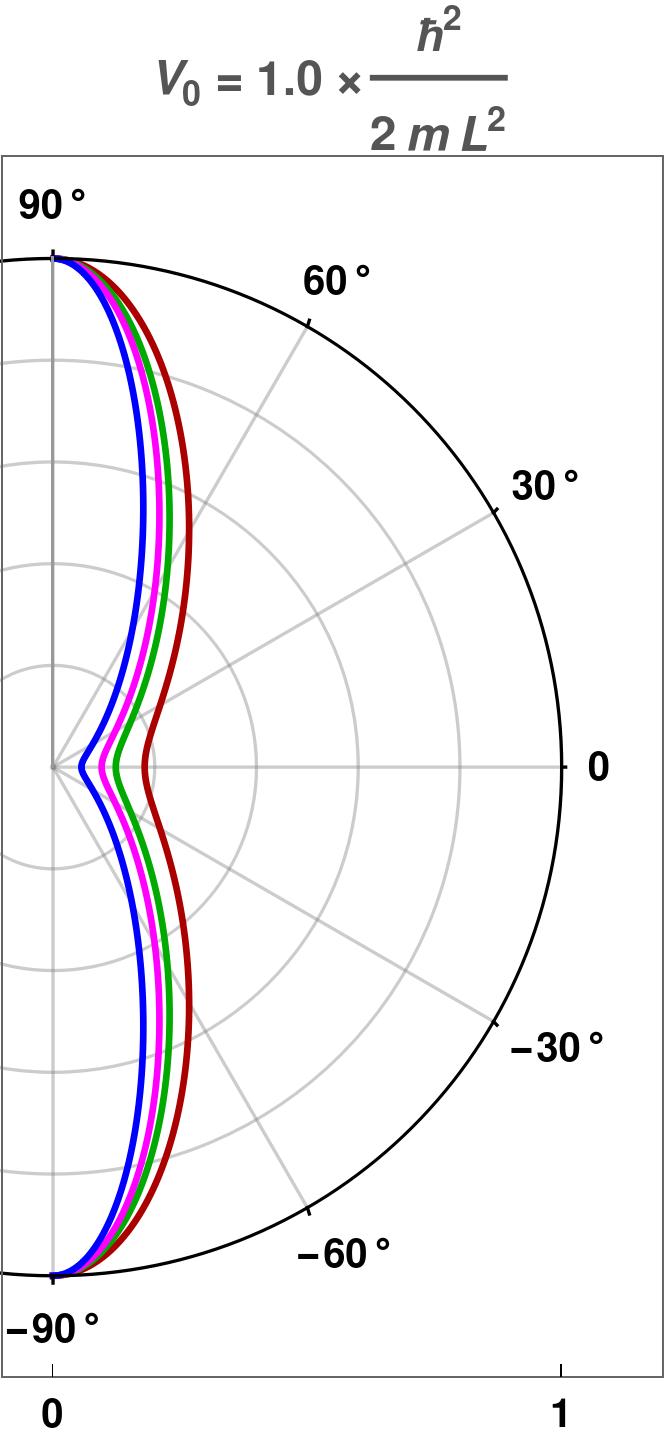}} \hspace{2 cm}
\subfigure[]{\includegraphics[width = 0.14 \textwidth]{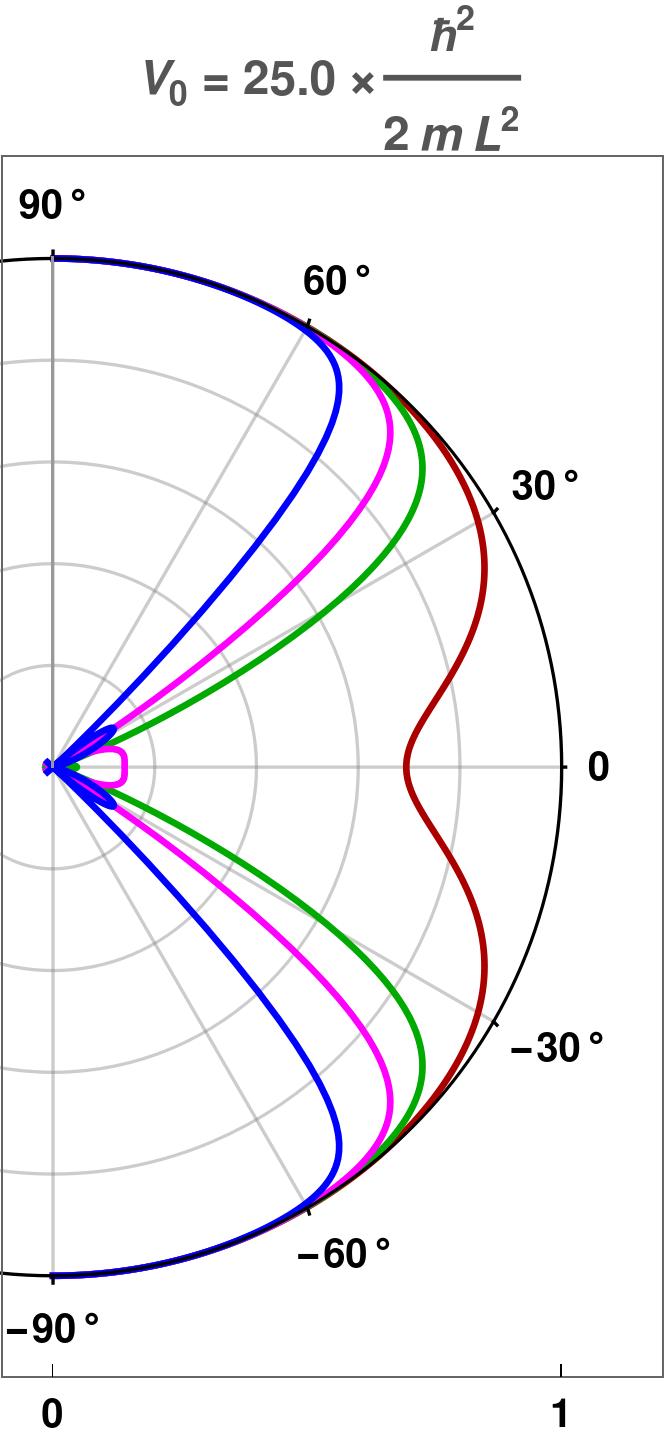}}\hspace{2 cm}
\subfigure[]{\includegraphics[width = 0.14 \textwidth]{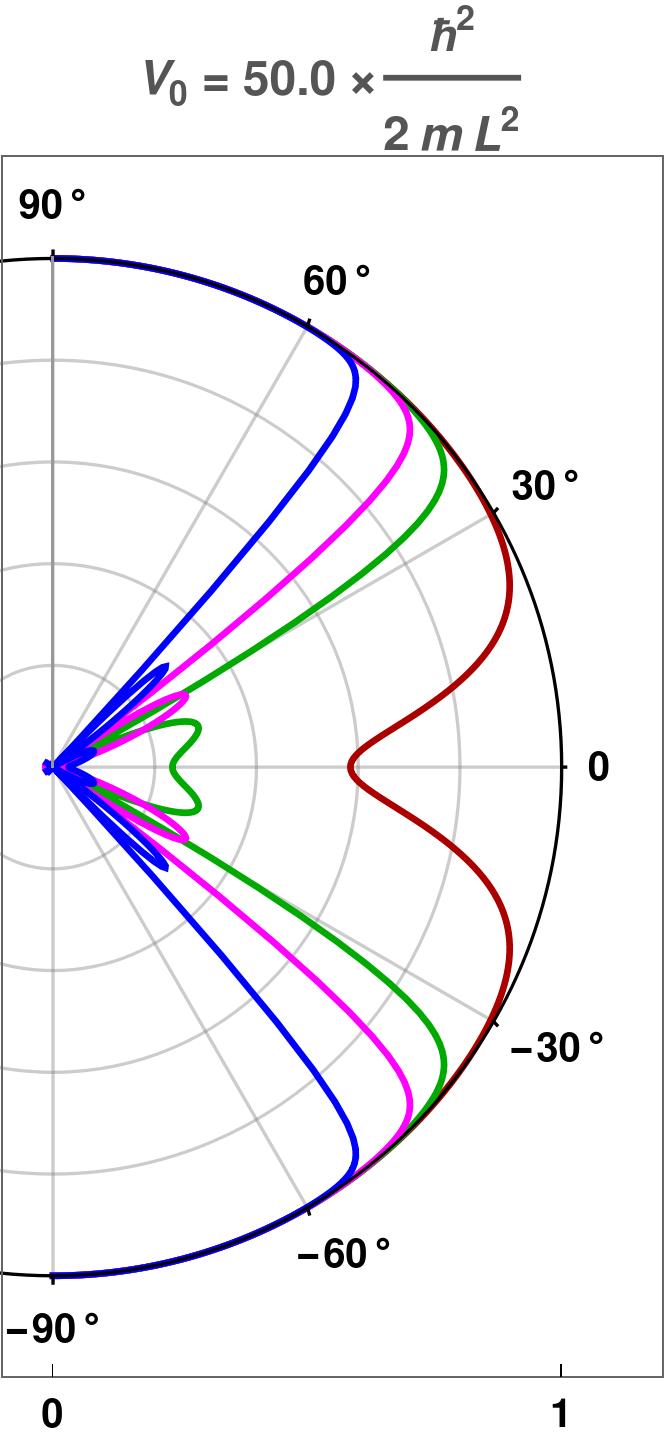}} \\
\subfigure[]{\includegraphics[width = 0.14 \textwidth]{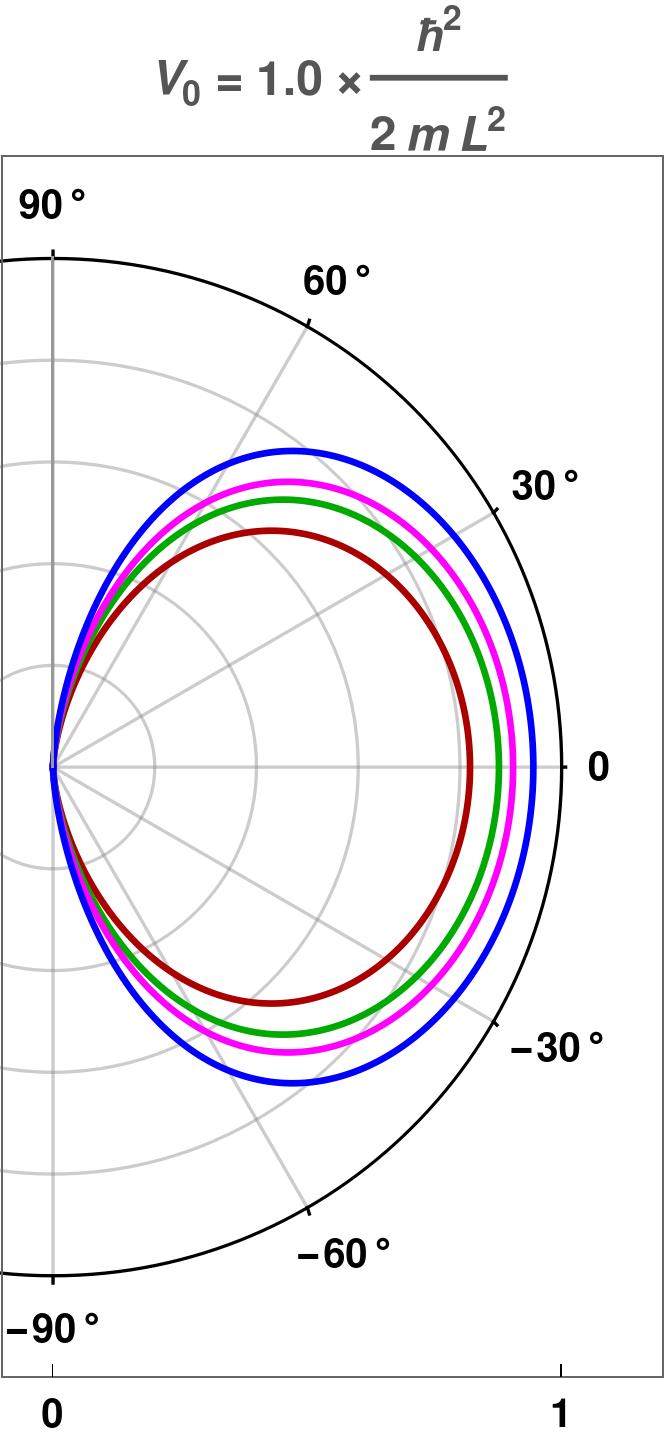}} \hspace{2 cm}
\subfigure[]{\includegraphics[width = 0.14 \textwidth]{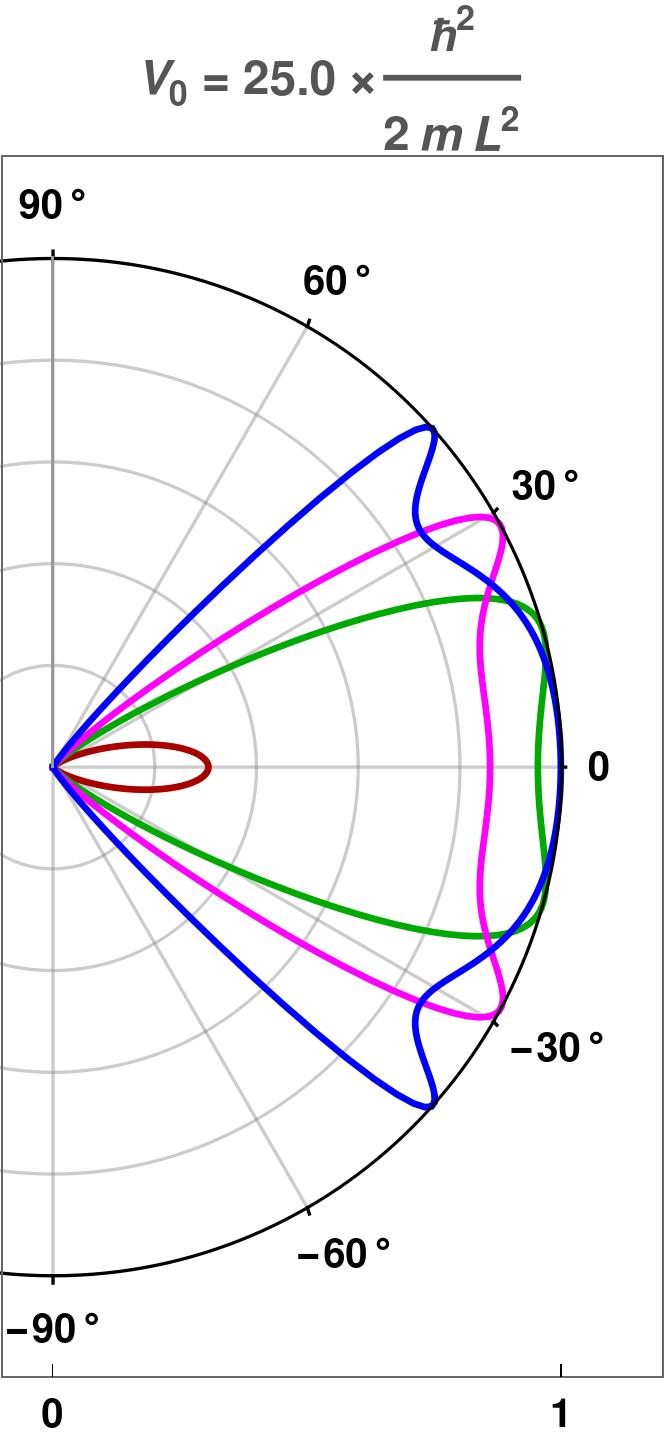}}\hspace{2 cm}
\subfigure[]{\includegraphics[width = 0.14 \textwidth]{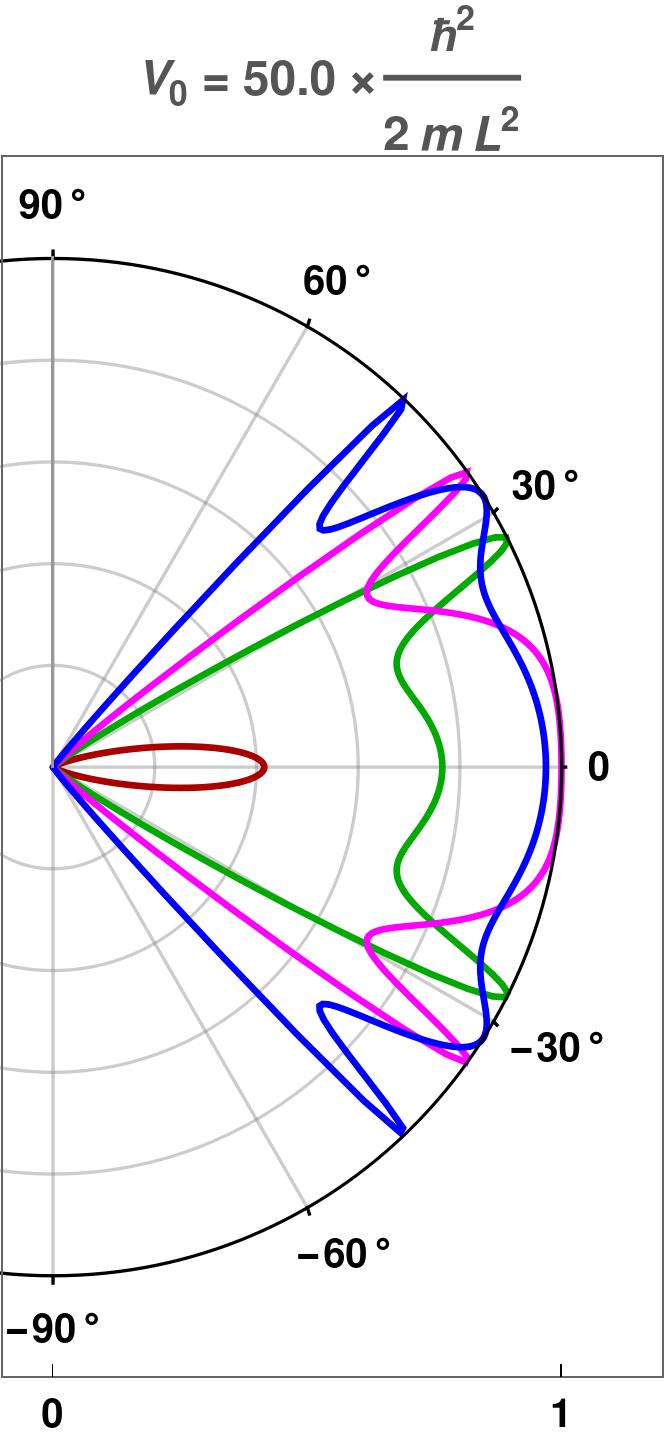}} 
\caption{2d QBCP: The polar plots show the reflection coefficient $R(E, V_0,\phi) \big \vert_{E>V_0}$ and the transmission coefficient $T(E, V_0,\phi) \big \vert_{E>V_0}$ as functions of the incident angle $\phi$ for the parameters $E=1.1\,V_0$ (red), $E=1.5\,V_0$ (green), $E=1.8\,V_0$ (magenta) and $E=2.5 \,V_0$ (blue).}
\label{figRTenMoreV}
\end{figure}
%%%%%%%%%%%%%%%%%%%%%%%

%%%%%%%%%%%%%%%%%%%%%%%%%%%%%%%%%%%%%%%%%%%%%%%%%%%
\begin{figure}[htb]
\subfigure[]
{\includegraphics[width = 0.45 \textwidth]{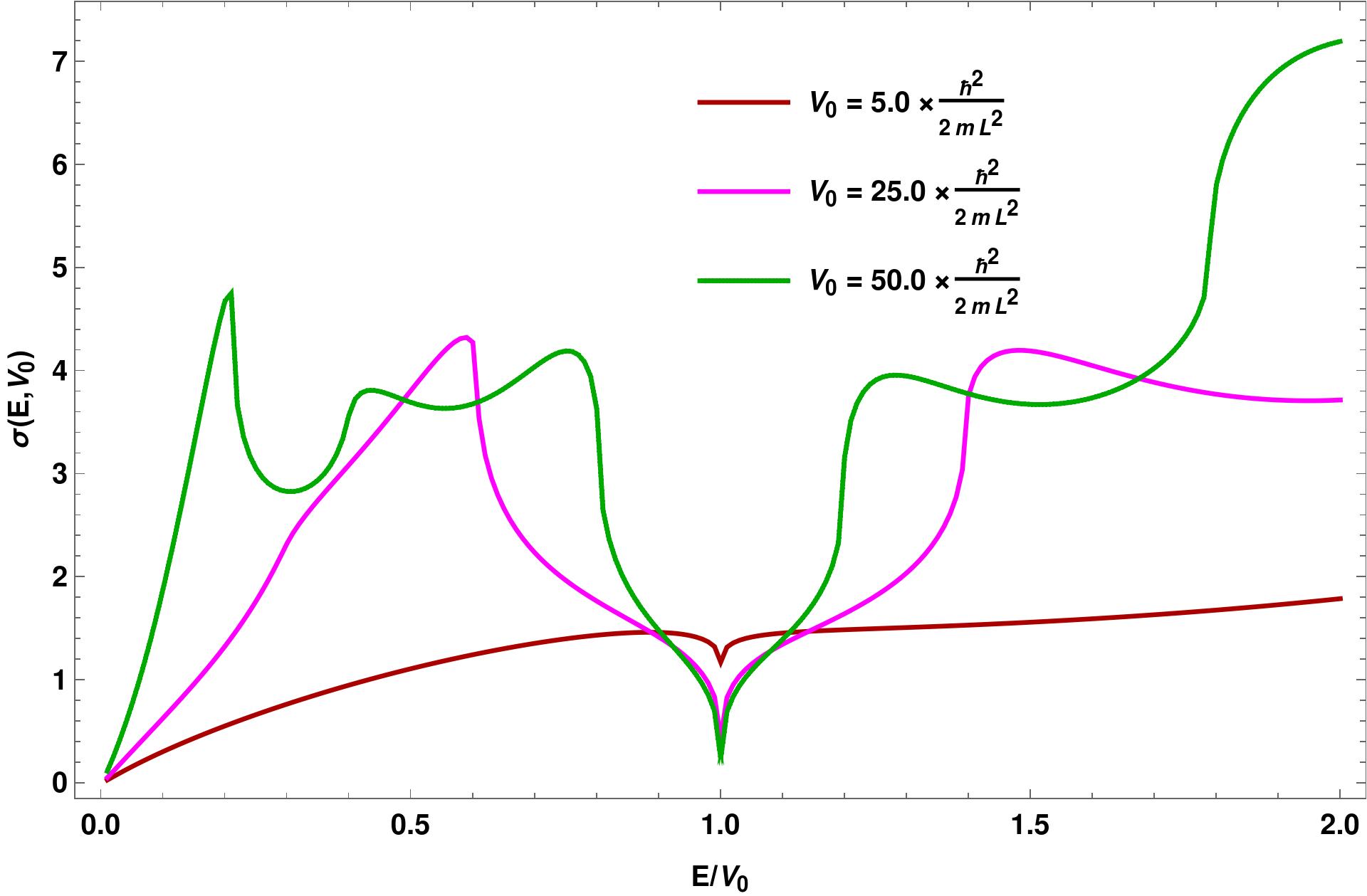}} \quad
\subfigure[]
{\includegraphics[width = 0.46 \textwidth]{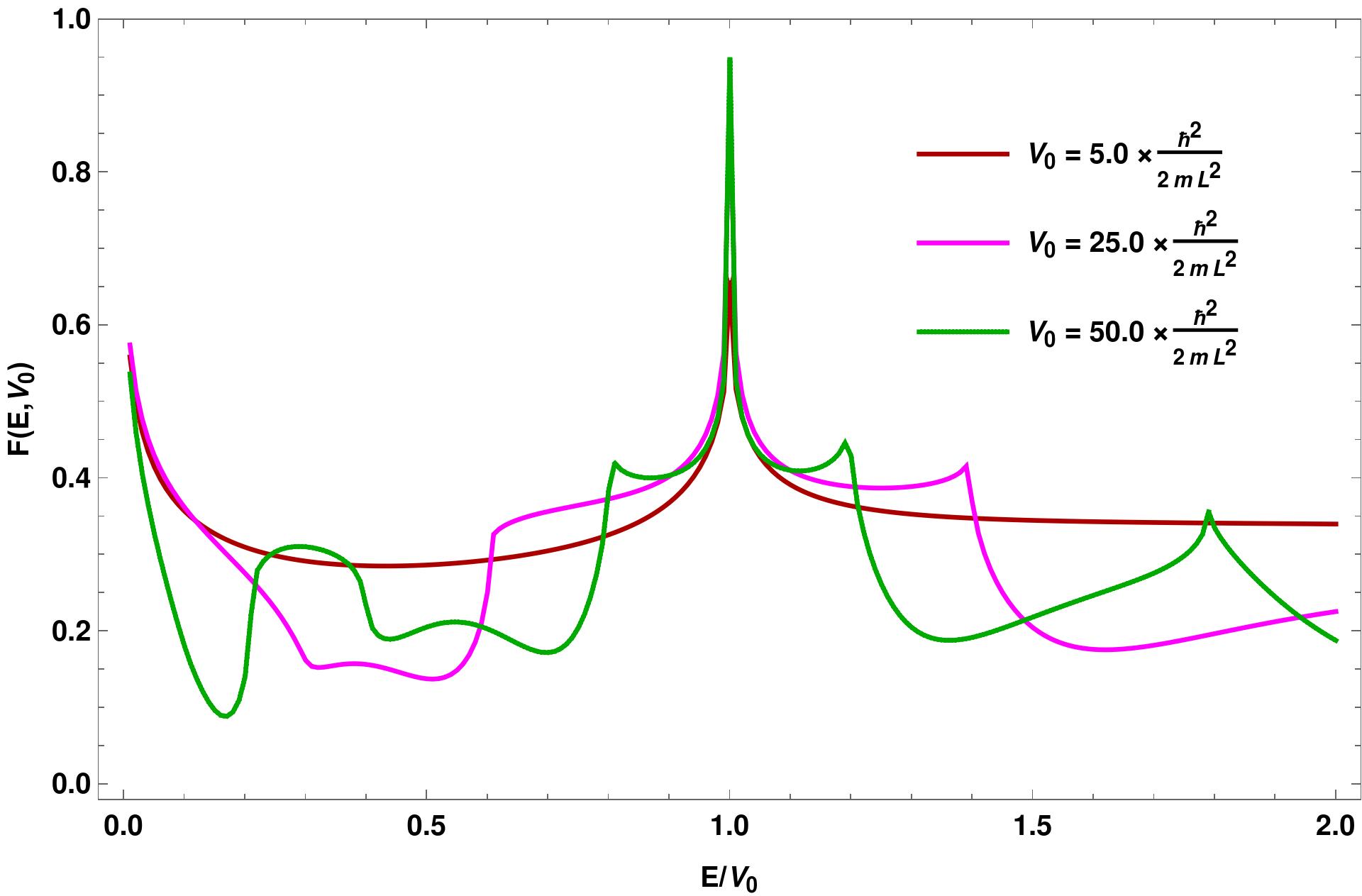}}
\caption{2d QBCP: Plots of the (a) conductivity ($\sigma$ in units of $2\pi$), and (b) Fano factor ($F$), as functions of $E/V_0$, for various values of $V_0$.}
\label{figfano}
\end{figure}
%%%%%%%%%%%%%%%%%%%%%%%

%%%%%%%%%%%%%%%%%%%%%%%%%%%%%%%%%%%%%%%%%%%%%%%
\subsection{Transmission coefficient, conductivity and Fano factor}
Let us assume $ W $ to be very large such that $q_n$ can effectively be treated as a continuous variable. We then numerically compute $T(E, \phi)$.

Using $ k_\ell = \sqrt{\frac{2mE} {\hbar^2}} \cos \phi\,,
\quad n=\frac{W\sqrt{ 2 m E }} {h}  \sin \phi\,,
\quad dn =\frac{2\,W\sqrt{ 2m E }} {h}  \cos \phi \, d\phi $, 
in the zero-temperature limit  and  for  a  small  applied  voltage,  the conductance is given by \cite{blanter-buttiker}:
\begin{align}
G(E,V_0) & = \frac{e^2}{h} \sum_n   |t_n|^2 
\rightarrow \frac{ e^2} {h} \int |t_n( E )|^2 \,dn
= \frac{e^2\, W\sqrt{ 2\,m\,E }} {h^2}  \int_{-\frac{\pi}{2}}^{\frac{\pi}{2}}
T( E , V_0,\phi) \cos\phi \,d\phi \,.
\end{align}
Therefore, the conductivity is given by:
\begin{align}
\sigma (E,V_0)  & =\frac{L }{W} \frac{G(E,V_0)}{e^2/h}
=  2\,\pi \sqrt{ \frac{ E}{\hbar^2/\left( 2\,mL^2 \right)} }  
\int_{-\frac{\pi}{2}}^{\frac{\pi}{2}} T( E , V_0, \phi) \cos\phi \,d\phi\,.
\end{align}
Shot noise is the measure of the fluctuations of the current away from their average value. The zero-temperature shot noise
is given by \cite{blanter-buttiker}:
\begin{align}
 \mathcal{S} (E,V_0) & =\frac{2\,e^2\,\Phi}{h} \sum_n  |t_n|^2\,  |r_n|^2 
\rightarrow\frac{ e^3\,\Phi\, W \sqrt{ \frac{ E}{\hbar^2/\left( 2\,mL^2 \right)} }  }
 { \pi \,h \,L}  
\int_{-\frac{\pi}{2}}^{\frac{\pi}{2}}
T( E ,  V_0,\phi)  \left[1-T( E , V_0, \phi) \right] \,d\phi \,,
\end{align}
where $\Phi $ is the applied voltage, and is characterized by the Fano factor:
\begin{align}
F(E,V_0)  =\frac 
{\int_{-\frac{\pi}{2}}^{\frac{\pi}{2}}
T( E ,  V_0,\phi)  \left[1-T( E ,  V_0,\phi) \right] \,d\phi }
{\int_{-\frac{\pi}{2}}^{\frac{\pi}{2}}
T( E , V_0, \phi) \,d\phi }  \,.
\end{align}

We express $E$ and $V_0$ in units of $\frac{\hbar^2} {2 \,m L^2}$, and study the behaviour of $T( E , V_0,\phi)$, $\sigma (E,V_0)$ and $ F (E,V_0)$. Figs.~\ref{figRTenLessV} and \ref{figRTenMoreV} show the polar plots of $R( E , V_0,\phi)$ and $T( E , V_0,\phi)$ as functions of the incident angle $\phi$, for the cases $E\leq V_0$ and $E>V_0 $ respectively.
They also serve to demonstrate that $R( E , V_0,\phi)+T( E , V_0,\phi)=1$.
From the expression of transmission coefficient in Eq.~(\ref{eqtval}), it is clear that the transmission is zero at normal incidence $(\phi=0)$, as long as $E<V_0$. Hence, we do not have a Klein tunneling analogue in the 2d QBCP, unlike graphene \cite{geim} or three-band pseudospin-1 Dirac-Weyl systems \cite{fang,zhu}. However, we still have the resonance conditions $\tilde k \,L= \pi N,\, N\in \mathbb{Z}$, under which the barrier becomes transparent $(T= 1)$. In Fig.~\ref{figfano}, we illustrate the conductivity $\sigma (E,V_0)$ and the Fano factor $ F (E,V_0)$, as functions of $E/V_0$, for some values of $V_0$.

%%%%%%%%%%%%%%%%%%%%%%%%%%%%%%%%%%%%%%%%%%%%%%%%%%%%%%%%%%%%%%
\section{3d Model}
\label{sec3dmodel}

We consider a model for 3d QBCP semimetals, where the low energy bands form a four-dimensional representation of the lattice symmetry group \cite{MoonXuKimBalents}. Then the standard $\left(\mathbf{k} \cdot \mathbf{p} \right)$ Hamiltonian for the particle-hole symmetric system can be written by using the five $4\times 4$ Euclidean Dirac matrices $\Gamma_a$ as \cite{murakami,Herbut}:
 \begin{equation}
 \mathcal{H}_{3d}^{kin}(k_x,k_y,k_z) = \frac{\hbar^2}{2\,m}
 \sum_{a=1}^5 d_a(\mathbf {k}) \,  \,\Gamma_a   \,.
\label{bare}
 \end{equation}
%%%%%%%%%%%%%%%%
The $\Gamma_a$'s form one of the (two possible) irreducible, four-dimensional Hermitian representations of the five-component Clifford algebra defined by the anticommutator $\{ \,\Gamma_a, \,\Gamma_b \} = 2\, \delta_{ab}$.
The five anticommuting gamma-matrices can always be chosen such that three are real and two are imaginary \cite{murakami,igor12}. In the representation used here, $(\Gamma_1, \Gamma_3, \Gamma_5)$ are real and $(\Gamma_2, \Gamma_4 ) $ are imaginary:
\begin{align}
\Gamma_1 = \sigma_1 \otimes \sigma_0 \,, \quad  \Gamma_2 = \sigma_2 \otimes \sigma_0 \,,
\quad \Gamma_3 = \sigma_3 \otimes \sigma_1 \,, \quad \Gamma_4 = \sigma_3 \otimes \sigma_2 \,,\quad
\Gamma_5 = \sigma_3 \otimes \sigma_3 \,.
\end{align}
The five functions $  d_a(\mathbf{k})$ are the real $\ell=2$ spherical harmonics, with the following structure:
\begin{align}
& d_1(\mathbf k)= \sqrt{3}\,k_y\, k_z \, , \,\,  
d_2(\mathbf k)= \sqrt{3}\,k_x \,k_z \,, \,\, 
d_3(\mathbf k)= \sqrt{3}\,k_x\, k_y  , \nonumber \\
& d_4(\mathbf k) = \frac{\sqrt{3}}{2}(k_x^2 -k_y^2) \,, \,\, 
d_5(\mathbf k) =
\frac{1}{2}\left (2\, k_z^2 - k_x^2 -k_y^2 \right ) .
\end{align}
%%%%%%%%%%%%%%%%%%%%%
The energy eigenvalues are 
\begin{align}
\varepsilon_{3d}^{\pm}(k_x,k_y,k_z) = \pm  \frac{\hbar^2 \left( k_x^2 + k_y^2 +k_z^2 \right)}{2\,m} ,
\end{align}
where the ``$+$" and ``$-$" signs, as usual, refer to the conduction and valence bands. Each of these bands is doubly degenerate.

A set of orthogonal eigenvectors is given by:
\begin{align}
\Psi_{+,1}^T & = \frac{1} { \mathcal{N}_{+,1} } \left\{-\frac{(k_x+\mathrm{i}\, k_y) \left(k+k_z\right)}{(k_x-\mathrm{i}\, k_y)^2},
\frac{\mathrm{i}\, \left( k+3 \,k_z\right)}{\sqrt{3} \,(k_x-\mathrm{i}\, k_y)},
-\frac{\mathrm{i}\, \left(-2\, k_z \left(k+k_z\right)+k_x^2+k_y^2\right)}{\sqrt{3}\, (k_x-\mathrm{i}\, k_y)^2},1\right\} \,,
\nn
\Psi_{+,2}^T
&=  \frac{1} { \mathcal{N}_{+,2} } 
\left\{\frac{(k_x+\mathrm{i}\, k_y) \left(k-k_z\right)}{(k_x-\mathrm{i}\, k_y)^2},
-\frac{\mathrm{i}\, \left(k-3\, k_z\right)}{\sqrt{3} \,(k_x-\mathrm{i}\, k_y)},-\frac{\mathrm{i}\, \left(2\, k_z \left( k-k_z\right)+k_x^2+k_y^2\right)}{\sqrt{3}\, (k_x-\mathrm{i}\, k_y)^2},1\right\} \,,
\nn
%%%%%%%%%%%%%%%%%%%%%
\Psi_{-,1}^T &= \frac{1} { \mathcal{N}_{-,1} } 
\left\{-\frac{\mathrm{i}\, \left(k+k_z\right)}{\sqrt{3}\, (k_x-\mathrm{i}\, k_y)},
\frac{ k-k_z}{k_x+\mathrm{i}\, k_y},1,-\frac{\mathrm{i}\, \left(2\, k_z \left(k_z-k\right)+k_x^2+k_y^2\right)}
{\sqrt{3}\, (k_x+\mathrm{i}\, k_y)^2}\right\} \,,\nn
%%%%%%%%%%%%%%%%%%%%
\Psi_{-,2}^T &= \frac{1} { \mathcal{N}_{-,2} } 
\left\{\frac{\mathrm{i}\, \left(k-k_z\right)}{\sqrt{3}\, (k_x-\mathrm{i}\, k_y)},
-\frac{k+k_z}{k_x+\mathrm{i}\, k_y},1,-\frac{\mathrm{i}\, \left(2 \,k_z 
\left(k+k_z\right)+k_x^2+k_y^2\right)}{\sqrt{3}\, (k_x+\mathrm{i}\, k_y)^2}\right\}\,,
\end{align}
where $k= \sqrt{k_x^2  + k_y^2 +k_z^2}$, and the ``$+$" (``$-$") indicates an eigenvector corresponding to the positive (negative) eigenvalue. The symbols $\frac{1} { \mathcal{N}_{\pm,1}}$ and $\frac{1} { \mathcal{N}_{\pm,2}}$ denote the corresponding normalization factors.

The 3d system is modulated by a square electric potential barrier of
height $V_0 $ and width $L$, as described in Eq.~(\ref{eqpot}).
Here, we need to consider the total Hamiltonian:
\begin{align}
\mathcal{H}_{3d}^{tot} &=
 \mathcal{H}_{3d}^{kin}(-\mathrm{i}\,\partial_x, -\mathrm{i}\,\partial_y, -\mathrm{i}\,\partial_z)+V(x)
\end{align} 
in position space. As before, we choose the $x$-axis along the transport direction, and place the chemical potential at an energy $E >0$ in the region outside the potential barrier. 

%%%%%%%%%%%%%%%%%%%%%%%%%%%%%%%%%%%%%%%%
\subsection{Formalism}

We consider the tunneling in a slab of height and width $W$.
Again, we assume that the material has a sufficiently large width $W$ along each of the two transverse directions, such that the boundary conditions are irrelevant for the bulk response, and impose the periodic boundary conditions:
\begin{align}
 \tilde \Psi^{\mathrm{tot}}(x,0,z) =
 \tilde \Psi^{\mathrm{tot}}(x,W,z)\,,\quad
\tilde \Psi^{\mathrm{tot}}(x,y,0) = 
\tilde \Psi^{\mathrm{tot}}(x,y,W) \,.
\end{align}
The  transverse momentum $ \mathbf k_\perp=(k_y,k_z)$ is conserved, and its components are quantized. Due to periodicity, we conclude that:
\begin{align}
k_y=\frac{ 2\,\pi\,n_y} {W} \equiv q_{n_y} \,,\quad k_z =\frac {2\,\pi\,n_z} {W} \equiv q_{n_z}\,,
\end{align}
where $(n_x, n_y) \in \mathbb{Z}$.
For the longitudinal direction (along the $x$-axis), we seek plane wave solutions of the form $ e^{\mathrm{i}\,k_x x} $. Then the full wavefunction is given by:
\begin{align}
\tilde  \Psi^{\mathrm{tot}} (x,y,z,\mathbf n) =
\text{const.}
\times \tilde  \Psi_{ \mathbf{n}}(x)\,  e^{ \mathrm{i}\left( q_{n_y} y +  q_{n_z} z\right ) }
\text{ with }
\mathbf{n} =(n_y, n_z)\,.
\end{align}

For any mode of given transverse momentum component $\mathbf{k}_\perp $, we can determine the $x$-component of the wavevectors of the incoming, reflected, and transmitted waves (denoted by $k_{\ell}$), by solving $
\varepsilon_{3d}^\pm(k_x ,  \mathbf n) =  \pm \frac{ \hbar^2 \left( k_{\ell}^2 + {\mathbf k_{\perp}}^2  \right)} {2\,m}\,.$
In the regions $x<0$ and $x>L$, we have only propagating modes ($  k_\ell $ is real), while the $x$-components
in the scattering region (denoted by $ \tilde k  $),  are given by $\tilde k^2 = \frac{2\,m\,|E-V_0|} {\hbar^2} - {\mathbf k_{\perp}}^2 $,
and may be propagating ($\tilde k $ is real) or evanescent ($\tilde k $ is imaginary).

We will follow the same procedure as described for the 2d QBCP. Again, without any loss of generality, we consider the transport of one of the degenerate positive energy states ($\Psi_{+,1}$) corresponding to electron-like particles, with the Fermi level outside the potential barrier being adjusted to a value $E =\varepsilon_{3d}^+(k_x,k_y, k_z) $.
In this case, a scattering state $\tilde \Psi_{ \mathbf n}$, in the mode labeled by $\mathbf n$, is
constructed from the states:
\begin{align}
 \tilde \Psi_{ \mathbf n} (x)=&  \begin{cases} \tilde \phi_L & \text{ for } x<0 \,, \\
\tilde \phi_M & \text{ for } 0< x < L \,,\\
\tilde \phi_R &  \text{ for } x > L \,,
\end{cases} \nonumber \\
%%%%%%%%%%%%%%%%%%%%%%%%%%%%%%%%%%%%%%%%
 \tilde \phi_L = & \,\frac{   
 \Psi_{+,1} (   k_\ell, q_{n_y},q_{n_z}) \, e^{\mathrm{i}\, k_\ell x }
+  
\sum \limits_{s =1,2}
 r_{\mathbf n,s} \,\Psi_{+,s} ( -k_{\ell} ,q_{n_y},q_{n_z}) \, e^{-\mathrm{i}\, k_\ell x }
}
{\sqrt{ \tilde{ \mathcal{V} }(k_\ell, \mathbf n) }}
\, ,\nonumber \\
%%%%%%%%%%%%%%%%%%%%%%%%%%%%%%%%%%%%%%%%%%%%5
 \tilde \phi_M  = &\,\Big[  
 \sum \limits_{s =1,2}
 \alpha_{\mathbf n,s} \,\Psi_{+,s} ( \tilde k,q_{n_y},q_{n_z}) \,
 e^{\mathrm{i}\,\tilde k\,  x } 
 + 
\sum \limits_{s =1,2}
 \beta_{\mathbf n,s} \,\Psi_{+,s} (  -\tilde k,q_{n_y},q_{n_z}) \,
 e^{-\mathrm{i}\,\tilde k  \, x } 
\Big]   \Theta\left( E-V_0  \right)
  \nonumber\\
& 
 + \Big[
 \sum \limits_{s =1,2}
 \alpha_{\mathbf n,s} \,\Psi_{-,s} ( \tilde k,q_{n_y},q_{n_z}) \,
 e^{\mathrm{i}\,\tilde k\,  x } 
 + 
\sum \limits_{s =1,2}
 \beta_{\mathbf n,s} \,\Psi_{-,s} (  -\tilde k,q_{n_y},q_{n_z}) \,
 e^{-\mathrm{i}\,\tilde k  \, x }
   \Big]  \,\Theta\left( V_0-E  \right) \,,\nonumber \\
 %%%%%%%%%%%%%%%%%%%%%%%%%%%%%%%%%%%%%%%%%%%%%%%%%%%%%%%%%%%55
\tilde \phi_R = & \,\frac{ \sum \limits_{s =1,2}
 t_{\mathbf n,s} \,\Psi_{+,s} ( k_{\ell} ,q_{n_y},q_{n_z}) 
 } 
{\sqrt{ \tilde{ \mathcal{V} }  (k_\ell, \mathbf n)}}
\, e^{\mathrm{i}\, k_\ell \left( x-L\right)},\nonumber \\
%%%%%%%%%%%%%%%%%%%%%%%%%%%%%%%%%
\tilde{ \mathcal{V} }(k_\ell, \mathbf n) \equiv &  \, |\partial_{k_\ell} \varepsilon_{3d}^+ (k_\ell, \mathbf n)|
= \frac{ \hbar^2  k_\ell } {m}\,,
\quad k_{\ell} = \sqrt{\frac{2\,m\,E} {\hbar^2}-q_{n_y}^2- q_{n_z}^2}
\,,\quad  \tilde k = \sqrt{\frac{2\,m\,|E-V_0|} {\hbar^2} -q_{n_y}^2-q_{n_z}^2 }\,,
\end{align}
where we have used the velocity $ \tilde{ \mathcal{V} }(k_\ell, \mathbf n) $ to normalize the incident, reflected and transmitted plane waves. 
Note that for $V_0> E$, the Fermi level within the potential barrier lies within the valence band, and we must use the valence band wavefunctions in that regio

The usual mode-matching procedure at $x=0$ and $x=L$ gives us
\footnote{From wavefunction matching, we have two equations from the two boundaries. For 3d QPCB,
each of these equations has four components as each wavevector has four components.
Therefore we have eight equations for eight undetermined coefficients.
We do not need to match the first derivatives of the wavefunction as those will be redundant equations.}:
\begin{align}
r_{\mathbf n, 1}(E, V_0) & = \begin{cases}
\frac{ \sin \left ({\tilde k} L \right )
 \left (\mathrm{i}\,k_{y}   + k_{\ell} \right ) 
 \left [ k_{\ell}\, k_{z}+\mathrm{i}\, k_{y} \sqrt{k_{\ell}^2+k_{y}^2+k_{z}^2}\right) 
\left({\tilde k}^2 \left(8 k_{\ell}^2-k_{y}^2-k_{z}^2\right)+\left(k_{y}^2+k_{z}^2\right) \left(5 k_{\ell}^2-4 \left(k_{y}^2+k_{z}^2\right)\right)
\right ]}
 {2 \left ( k_{\ell}-\mathrm{i}\, k_{y} \right )^2 
 \sqrt{k_{\ell}^2+k_{y}^2+k_{z}^2}
 \, \left [ \sin \left ({\tilde k} L \right  ) 
\left \lbrace 
{\tilde k}^2 \left(4 k_{\ell}^2+k_{y}^2+k_{z}^2\right)+\left(k_{y}^2+k_{z}^2\right) \left(k_{\ell}^2+4 \left(k_{y}^2+k_{z}^2\right)\right)
\right \rbrace 
 -6\, \mathrm{i}\, {\tilde k} k_{\ell} \cos \left ({\tilde k} L \right  ) \left(k_{y}^2+k_{z}^2\right)
 \right ] } 
&\text{ for } E < V_0 \\
%%%%%%%%%%%%
\frac{\left(k_{\ell}^2-{\tilde k}^2\right) \sin \left ({\tilde k} L \right  ) (k_{\ell}+\mathrm{i}\, k_{y}) \left(k_{\ell} k_{z}+\mathrm{i}\, k_{y} \sqrt{k_{\ell}^2+k_{y}^2+k_{z}^2}\right)}{(k_{\ell}-\mathrm{i}\, k_{y})^2 \sqrt{k_{\ell}^2+k_{y}^2+k_{z}^2} \, \left [ \left({\tilde k}^2+k_{\ell}^2\right) 
\sin \left ({\tilde k} L \right  )
+2\, \mathrm{i}\, {\tilde k} k_{\ell} \cos \left ({\tilde k} L \right )\right ]}
&\text{ for } E > V_0
\end{cases} \,,\nonumber \\
%%%%%%%%%%%%%%%%%%%%%%%%%%%%%%%%%%%%%%%%%
%%%%%%%%%%%%%%%%%%%%%%%%%%%%%%%%%%%%%%%%%%%
r_{\mathbf n, 2}(E, V_0) & = \begin{cases}
-\frac{{\tilde k}^2 \left[ 4 k_{\ell}^2-5 \left(k_{y}^2+k_{z}^2\right)\right]
+\left(k_{y}^2+k_{z}^2\right) \left [ k_{\ell}^2-8 \left(k_{y}^2+k_{z}^2\right)\right]
 \sin \left( {\tilde k}  L \right)}
%%%  
{  \sin \left ({\tilde k} L \right  ) 
\left [ {\tilde k}^2 \left(4 k_{\ell}^2+k_{y}^2+k_{z}^2\right)+\left(k_{y}^2+k_{z}^2\right)
 \left \lbrace k_{\ell}^2+4 \left(k_{y}^2+k_{z}^2\right)\right \rbrace
\right]
-6\, \mathrm{i}\, {\tilde k} k_{\ell} \cos \left ({\tilde k} L \right  ) 
\left(k_{y}^2+k_{z}^2\right )} &
\\ \times  \frac{ 
k_{\ell} \left (k_{\ell}+\mathrm{i}\, k_{y} \right) 
 \left(\sqrt{k_{\ell}^2+k_{y}^2+k_{z}^2}+k_{z}\right)
\sqrt{ \left(k_{z} \left(k_{z}-\sqrt{k_{\ell}^2+k_{y}^2+k_{z}^2}\right)+k_{\ell}^2+k_{y}^2\right)}} 
{ \left( k_{\ell}-\mathrm{i}\, k_{y} \right )^2 
\sqrt{  k_{\ell}^2+k_{y}^2+k_{z}^2  } \,
\sqrt{ k_{z} \left(\sqrt{k_{\ell}^2+k_{y}^2+k_{z}^2}+k_{z}\right)+k_{\ell}^2+k_{y}^2 }} 
%%%%
&\text{ for } E < V_0 \\
%%%%%%%%%%%%
\frac{k_{\ell} \left ({\tilde k}^2-k_{\ell}^2 \right ) 
\sin \left ({\tilde k} L \right  ) (k_{\ell}+\mathrm{i}\, k_{y}) 
\left(\sqrt{k_{\ell}^2+k_{y}^2+k_{z}^2}+k_{z}\right) 
\sqrt{  k_{z} \left(k_{z}-\sqrt{k_{\ell}^2+k_{y}^2+k_{z}^2}\right)+k_{\ell}^2+k_{y}^2}}
{(k_{\ell}-\mathrm{i}\, k_{y})^2 \sqrt{k_{\ell}^2+k_{y}^2+k_{z}^2}
\, \sqrt{ k_{z} \left(\sqrt{k_{\ell}^2+k_{y}^2+k_{z}^2}+k_{z}\right)+k_{\ell}^2+k_{y}^2 } 
\, \left [ \left({\tilde k}^2+k_{\ell}^2\right) \sin \left ({\tilde k} L \right  )+2 \mathrm{i}\, {\tilde k} k_{\ell} \cos \left ({\tilde k} L \right  )\right ]}
&\text{ for } E > V_0 
\end{cases}\,,
\label{eqrval3d}
\end{align}
and
%%%%%%%%%%%%%%%%%%%%%%%%
\begin{align}
t_{\mathbf n, 1}(E, V_0) & = \begin{cases}
-\frac{6 \,\mathrm{i}\, {\tilde k} \,k_{\ell} \left(k_{y}^2+k_{z}^2\right)}
{\sin \left ({\tilde k} L \right  ) \left [ {\tilde k}^2 
\left(4\, k_{\ell}^2+k_{y}^2+k_{z}^2\right)
+\left ( k_{y}^2+k_{z}^2\right) \left \lbrace k_{\ell}^2+4 \left(k_{y}^2+k_{z}^2\right)\right \rbrace
\right ]
-6\, \mathrm{i}\, {\tilde k}\, k_{\ell} \cos \left ({\tilde k} L \right  ) \left(k_{y}^2+k_{z}^2\right)}
&\text{ for } E < V_0 \\
%%%%%%%%%%%%
\frac{ -2 \,\mathrm{i}\, {\tilde k} \,k_{\ell}}{\left({\tilde k}^2+k_{\ell}^2\right) \sin \left ({\tilde k} L \right  )+2 \, \mathrm{i}\, {\tilde k} \,k_{\ell} \cos \left ({\tilde k} L \right  )}
&\text{ for } E > V_0
\end{cases} \,, \nonumber \\
t_{\mathbf n, 2}(E, V_0) & =0\,.
\label{eqtval3d}
\end{align}
The reflection and transmission coefficients at an energy $E$ are given by 
\begin{align}
R( E ,  V_0,\theta , \phi) = | r_{\mathbf n, 1}( E, V_0 )|^2 +| r_{\mathbf n, 2}( E, V_0 )|^2
 \text{ and }
T( E ,  V_0,\theta , \phi) = | t_{\mathbf n, 1}( E, V_0 )|^2 
\end{align}
respectively,
where $\theta = \cos^{-1} \left( \frac{\hbar \,q_{n_z}} {\sqrt {2\, m\, E}} \right)$ and
$\phi = \tan^{-1} \left( \frac {q_{n_y}} {k_\ell} \right)$
define the incident angle (solid) of the incoming wave in 3d.

%%%%%%%%%%%%%%%%%%%%%%%%%%%%%%%%%%%%%%%%%%%%%%%%%%%
\begin{figure}[]
%%%%%%%%%%%%%%%%%%%%%
\subfigure[]
{\includegraphics[width = 0.3 \textwidth]{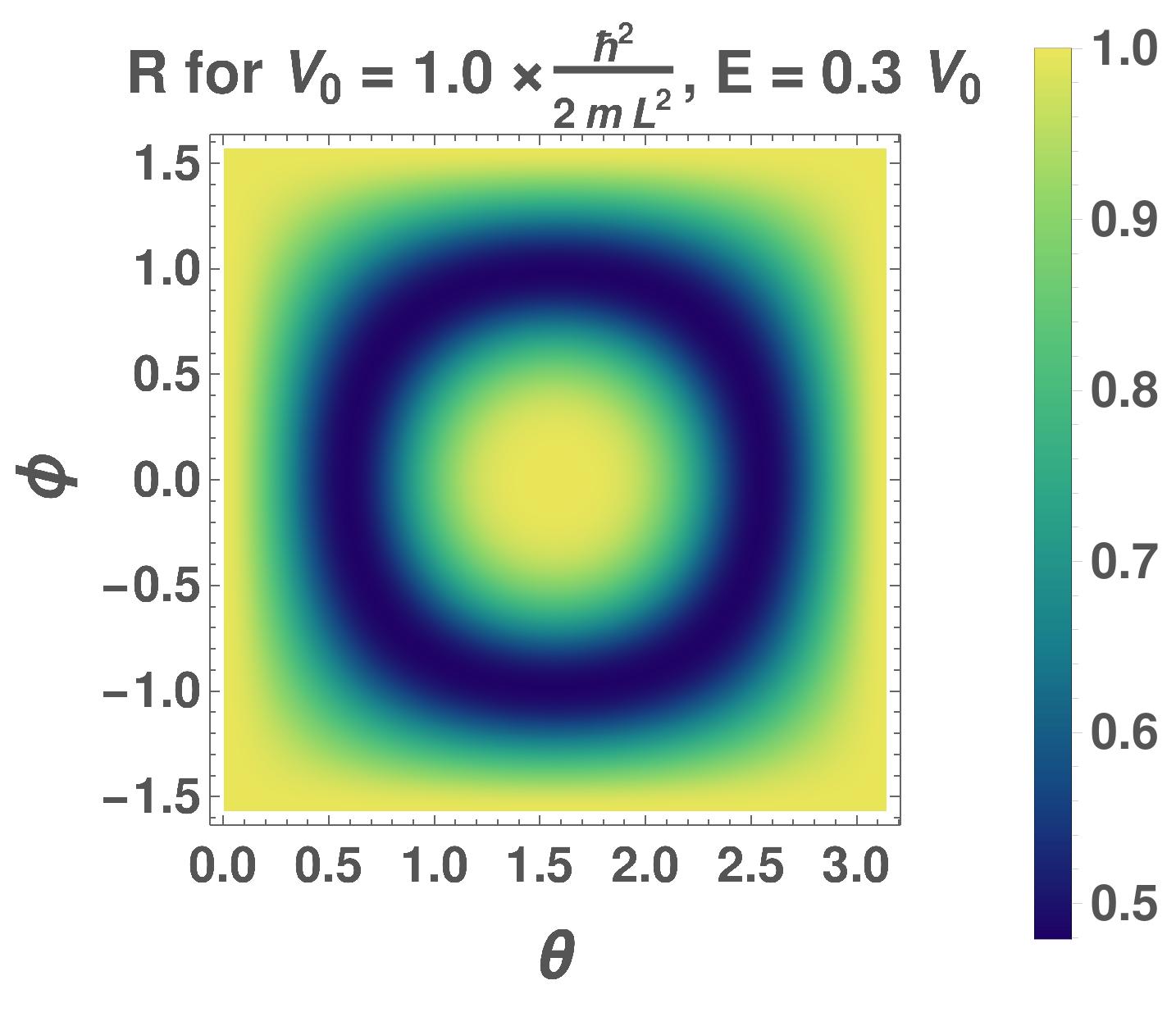}} \quad
\subfigure[]
{\includegraphics[width = 0.3 \textwidth]{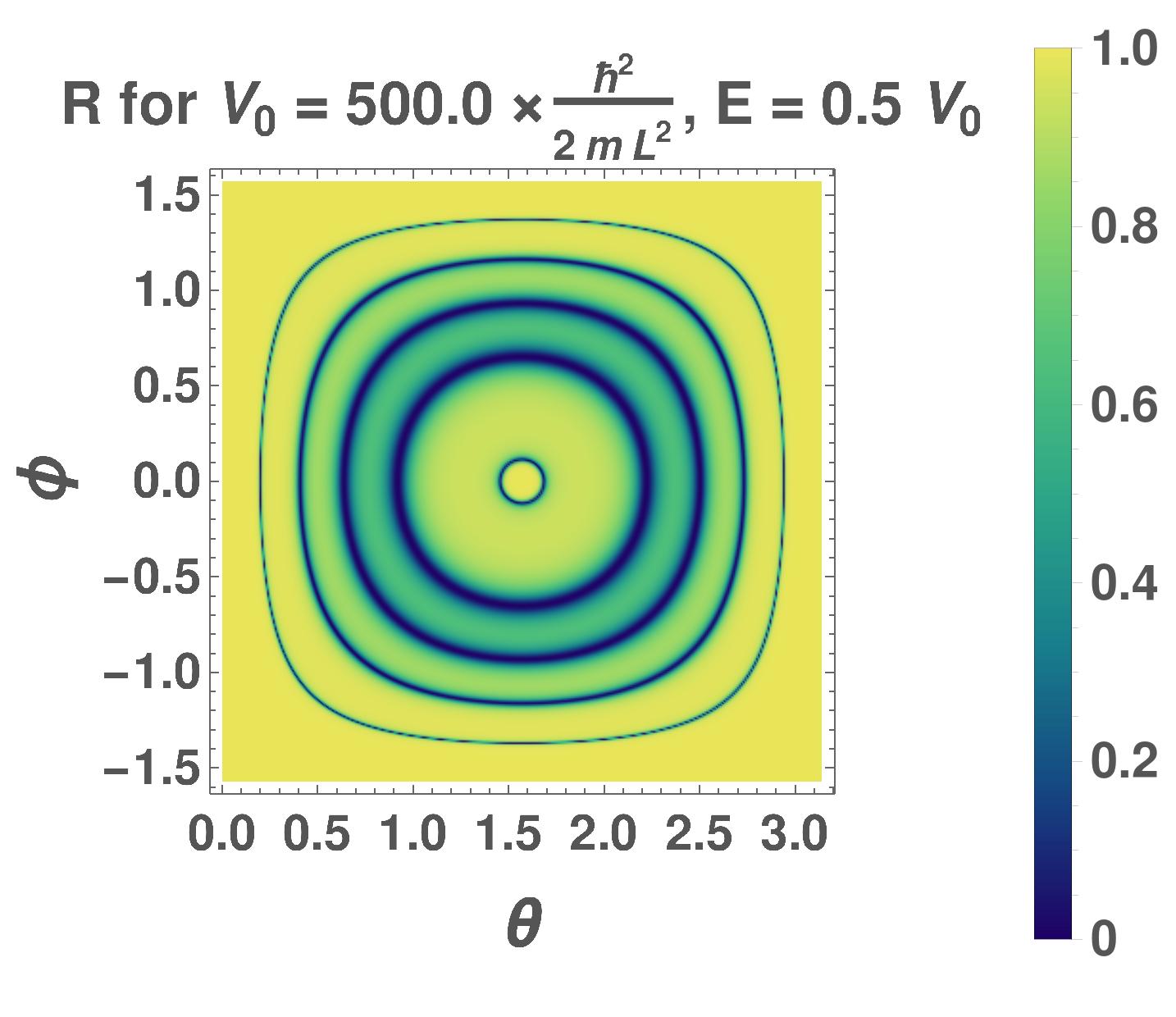}} \quad
\subfigure[]
{\includegraphics[width = 0.3 \textwidth]{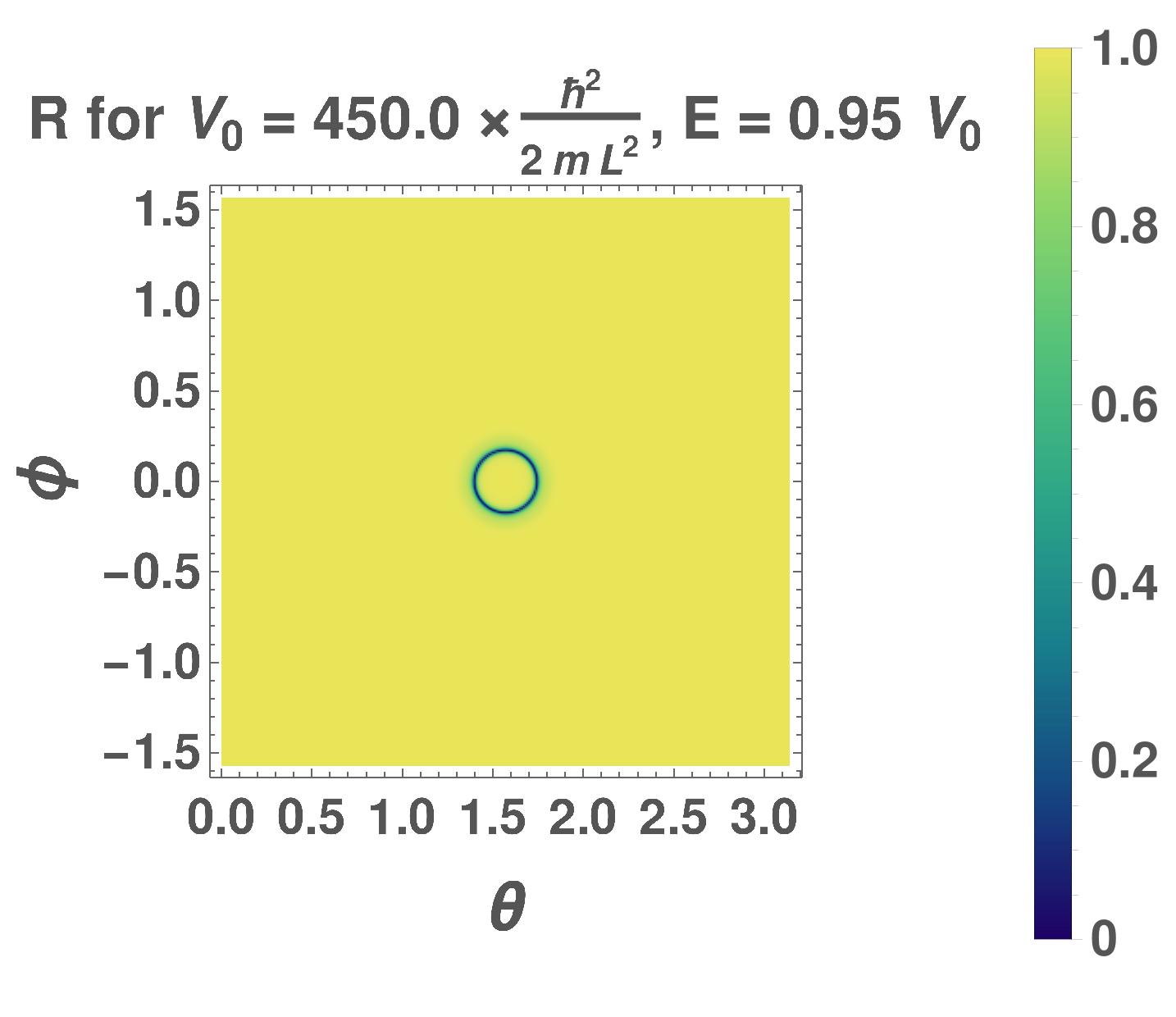}} \quad
%%%%%%%%%%%%%%%%%%%%%%%%%
\subfigure[]
{\includegraphics[width = 0.3 \textwidth]{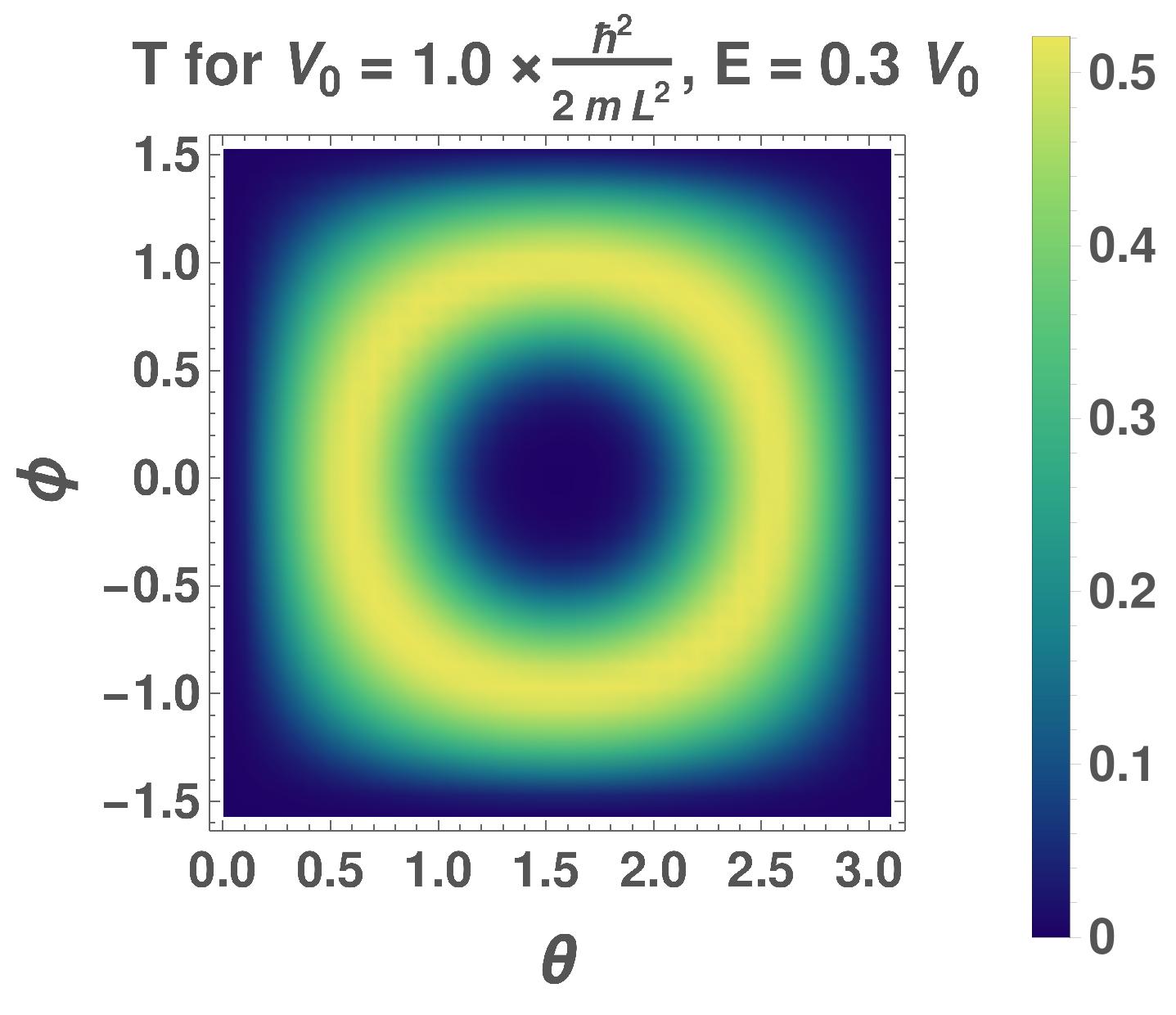}} \quad
\subfigure[]
{\includegraphics[width = 0.3 \textwidth]{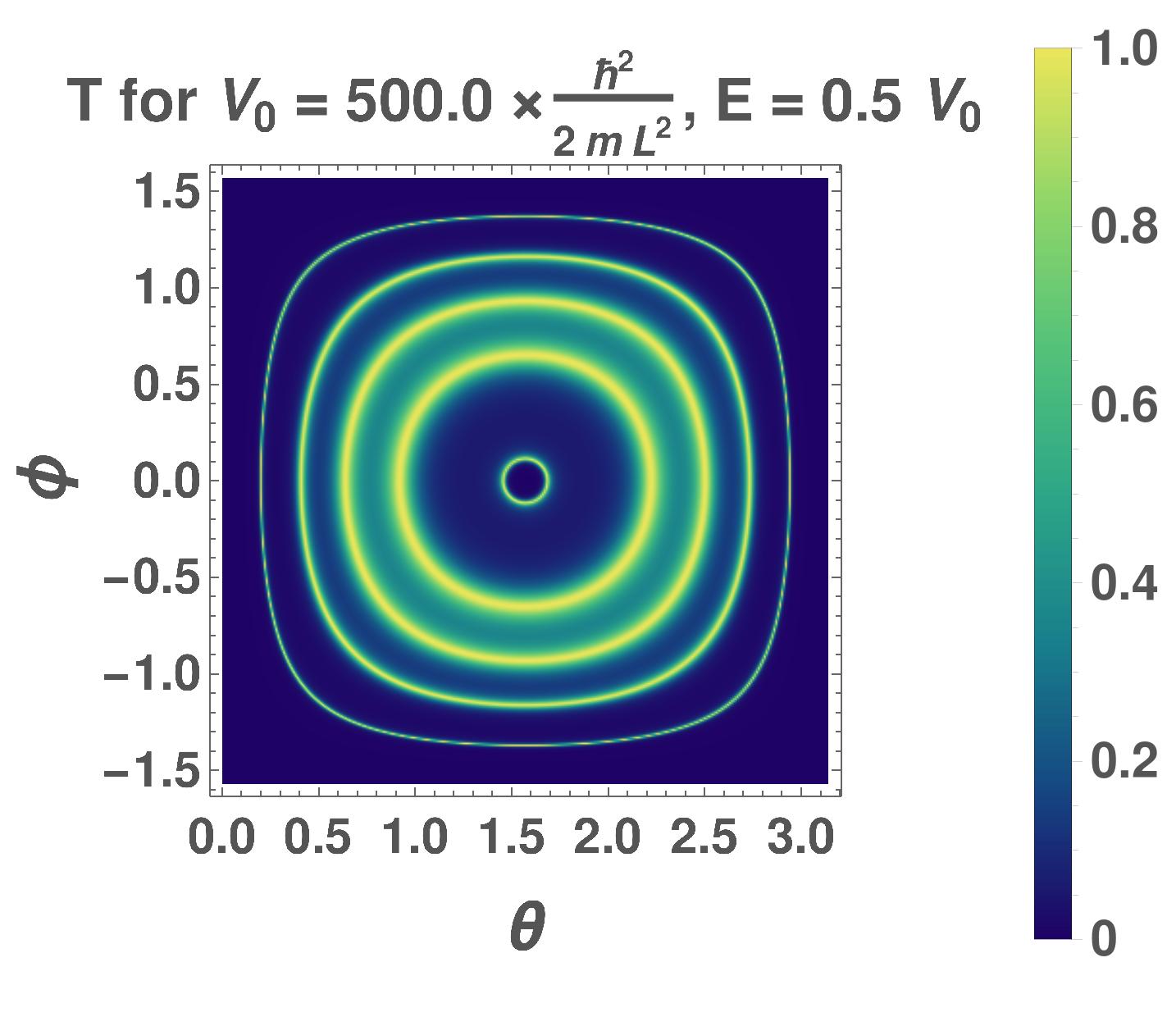}} \quad
\subfigure[]
{\includegraphics[width = 0.3 \textwidth]{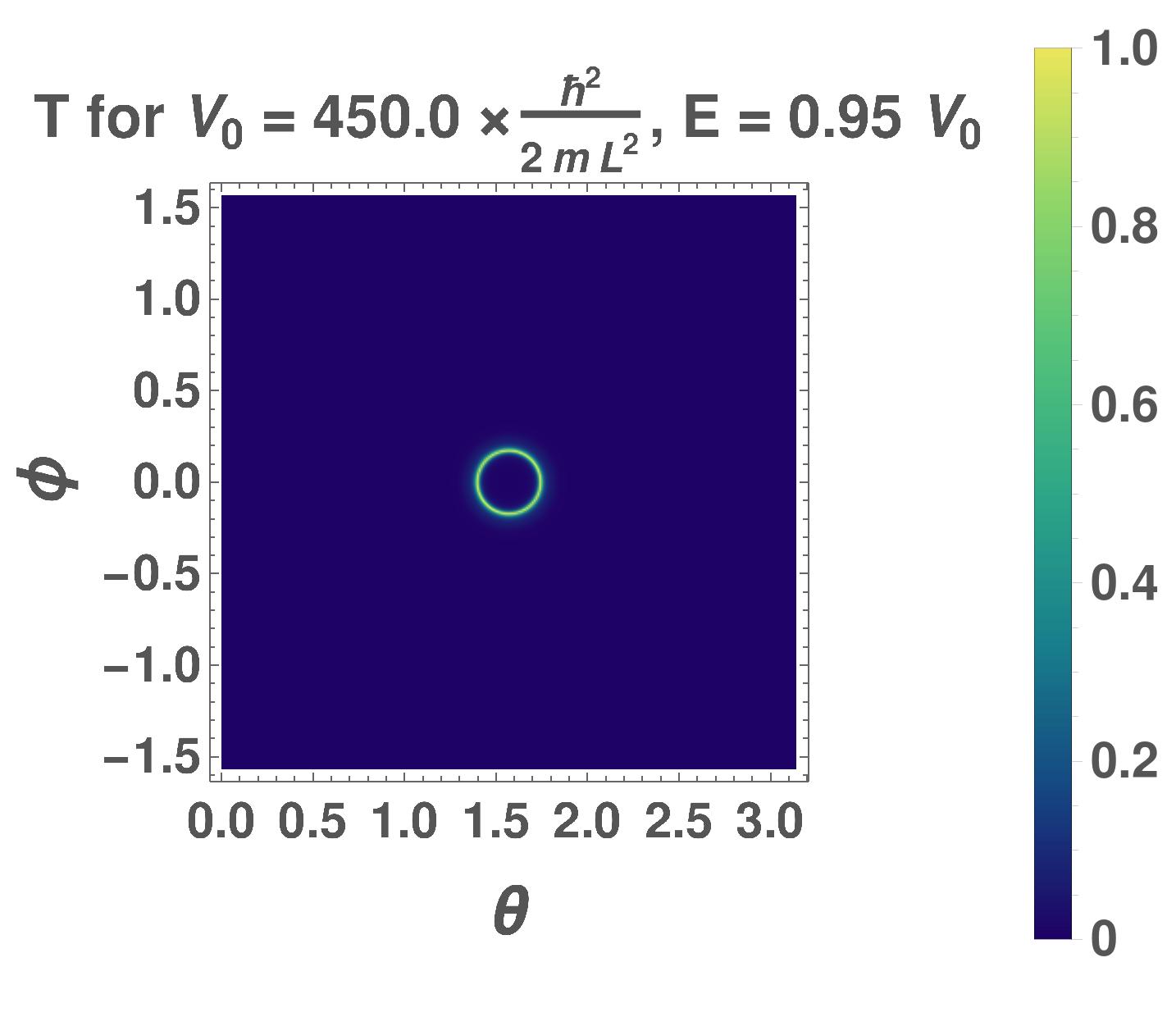}} \quad
%%%%%%%%%%%%%%%%%%%%%
\subfigure[]
{\includegraphics[width = 0.3 \textwidth]{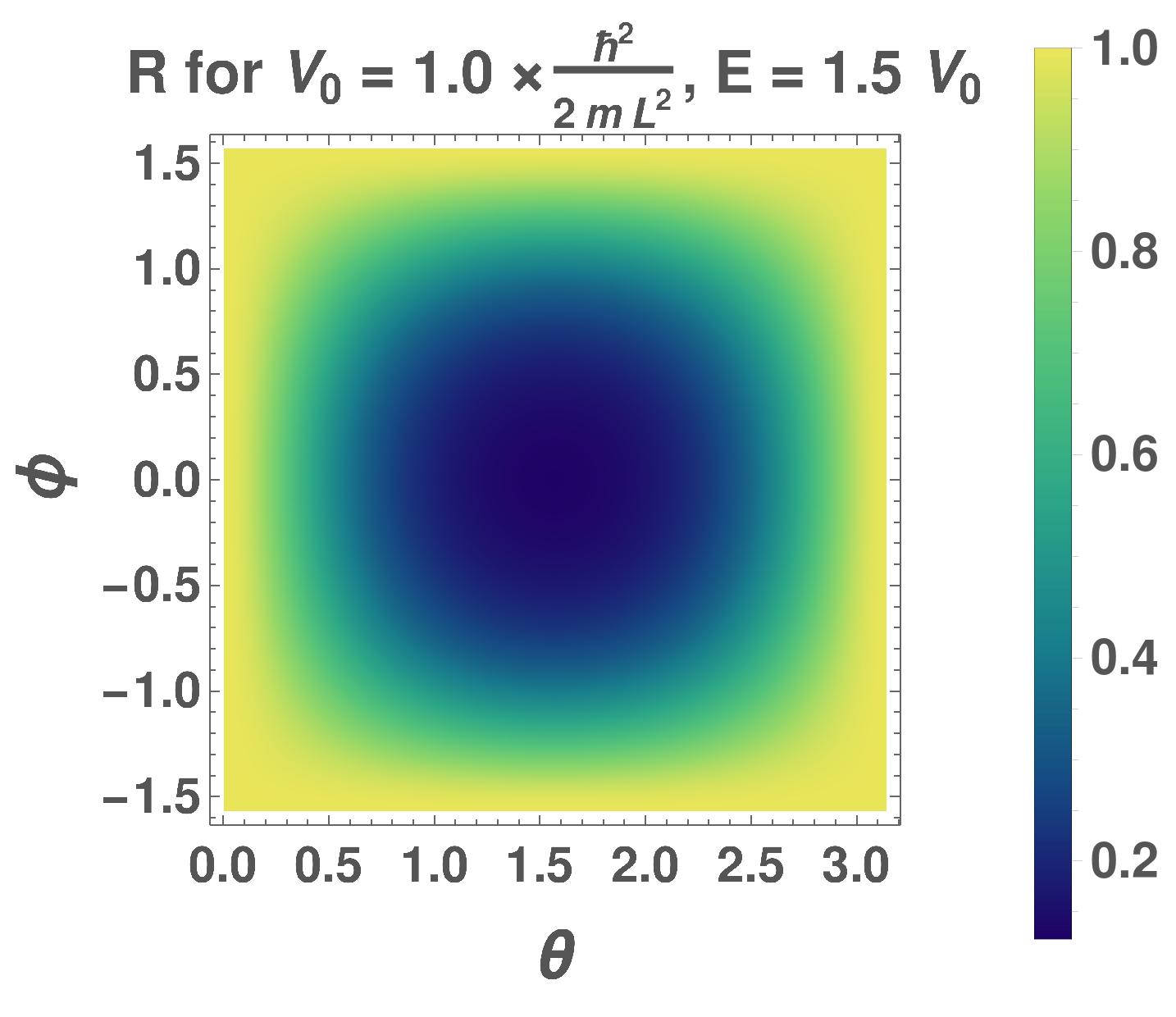}} \quad
\subfigure[]
{\includegraphics[width = 0.3 \textwidth]{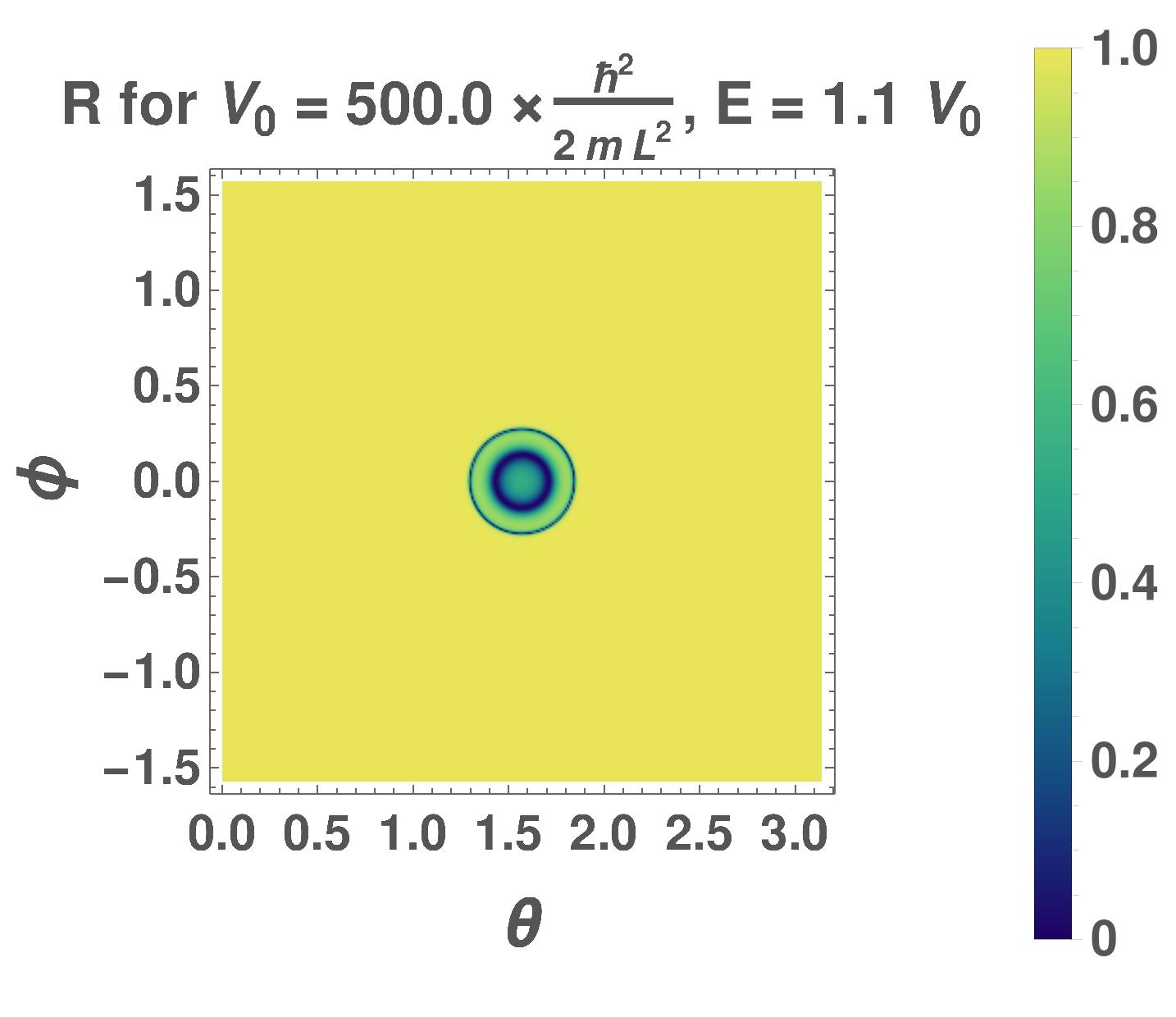}} \quad
\subfigure[]
{\includegraphics[width = 0.3 \textwidth]{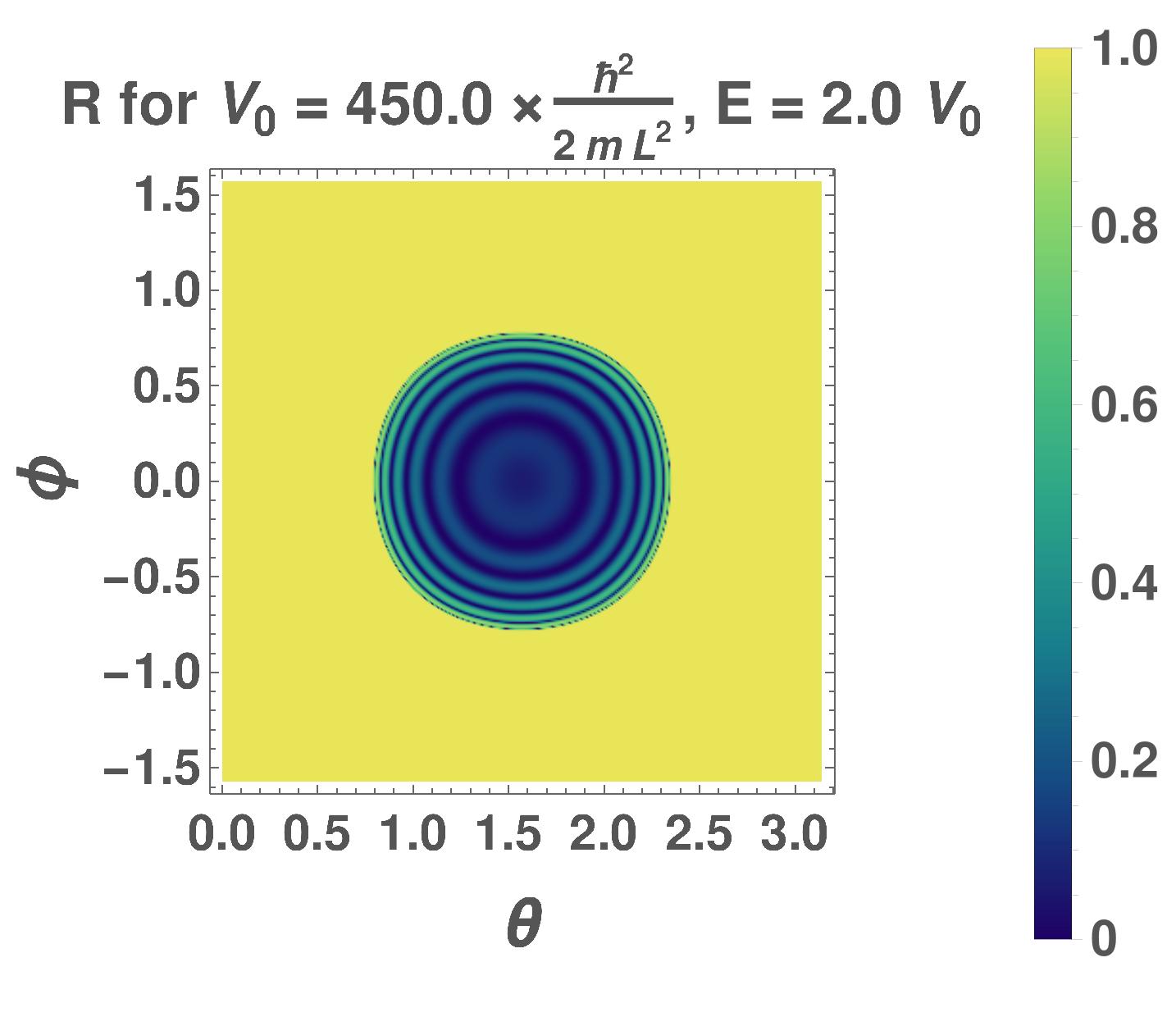}} \quad
%%%%%%%%%%%%%%%%%%%%%
\subfigure[]
{\includegraphics[width = 0.3 \textwidth]{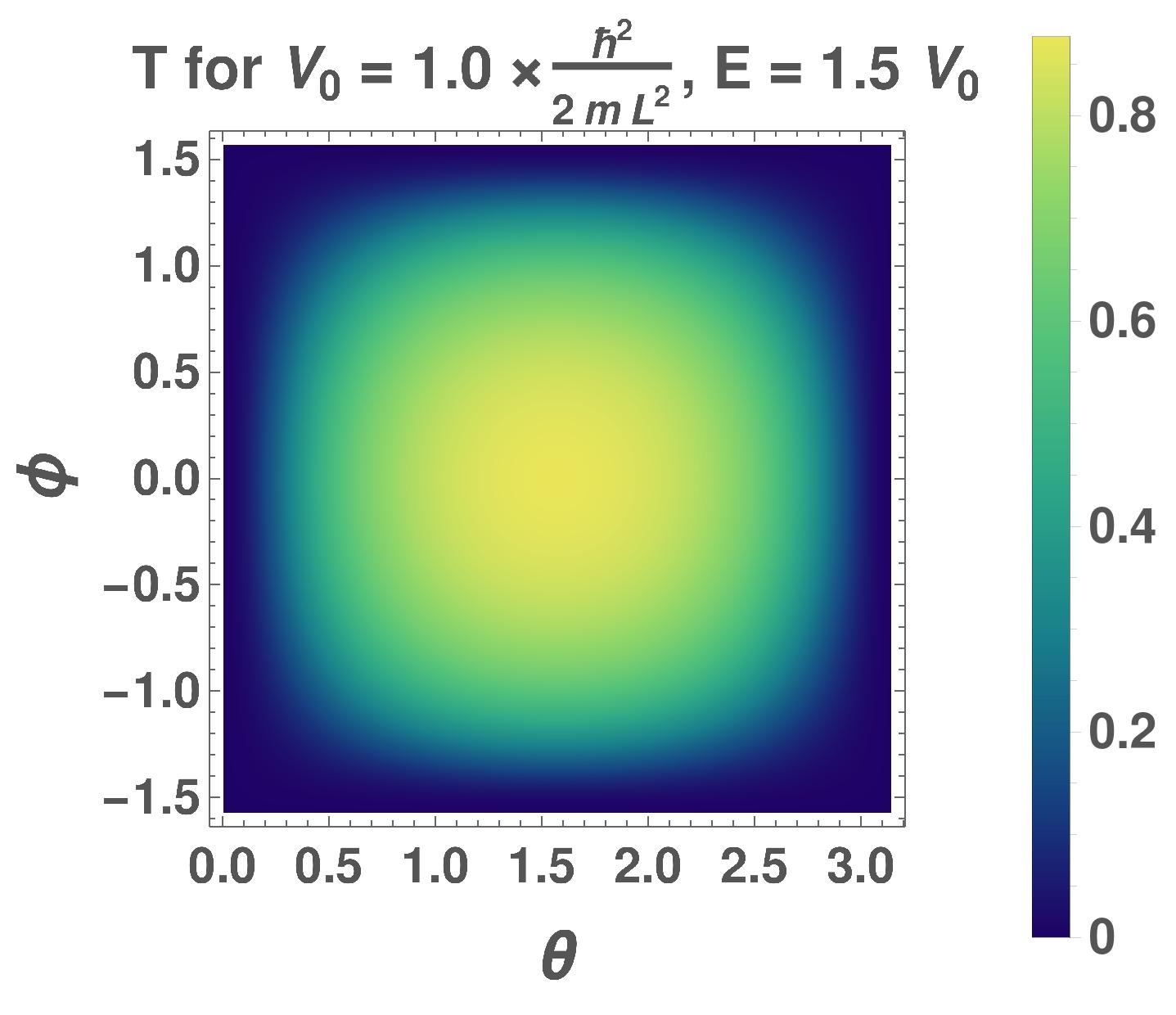}} \quad
\subfigure[]
{\includegraphics[width = 0.3 \textwidth]{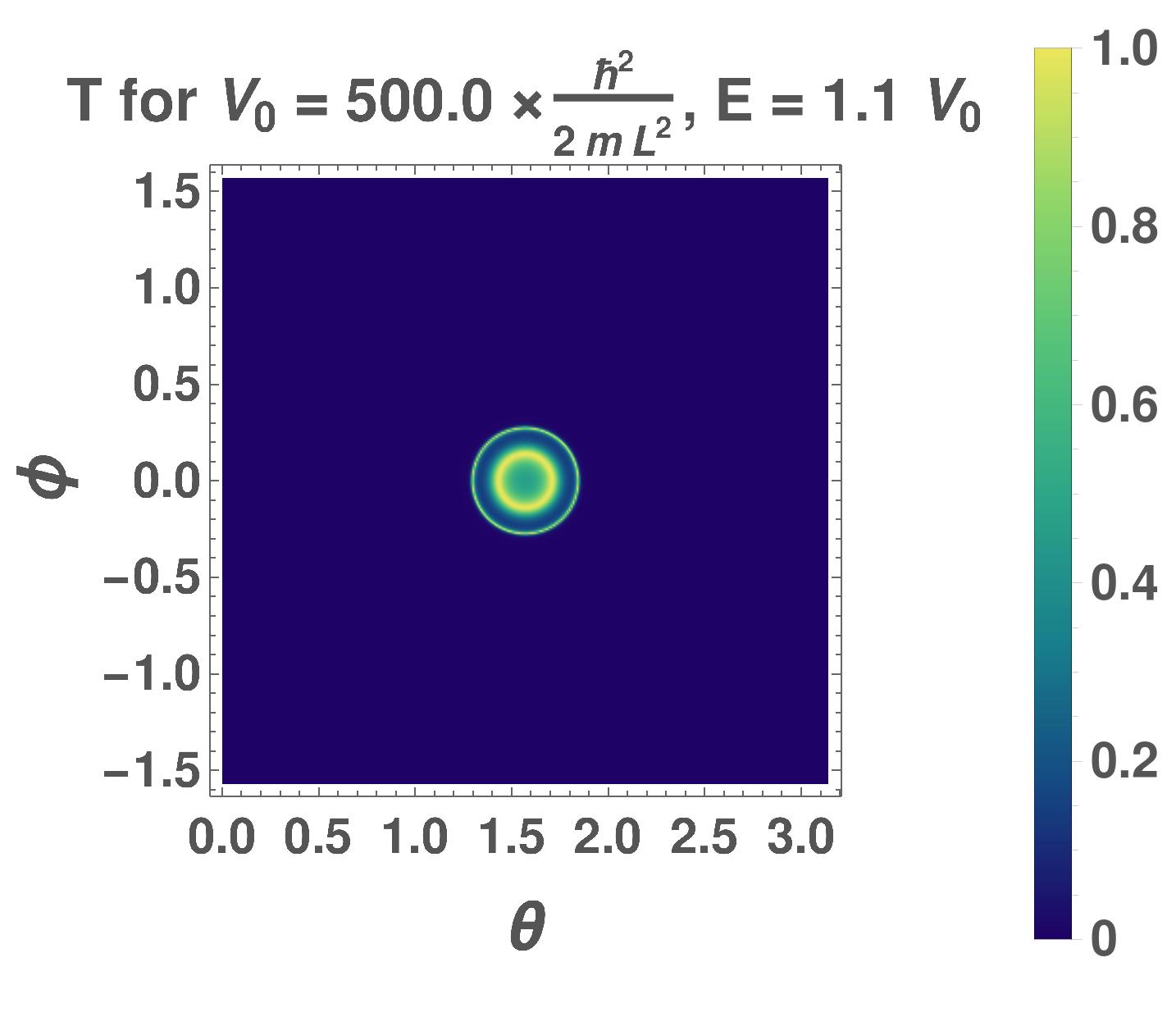}} \quad
\subfigure[]
{\includegraphics[width = 0.3 \textwidth]{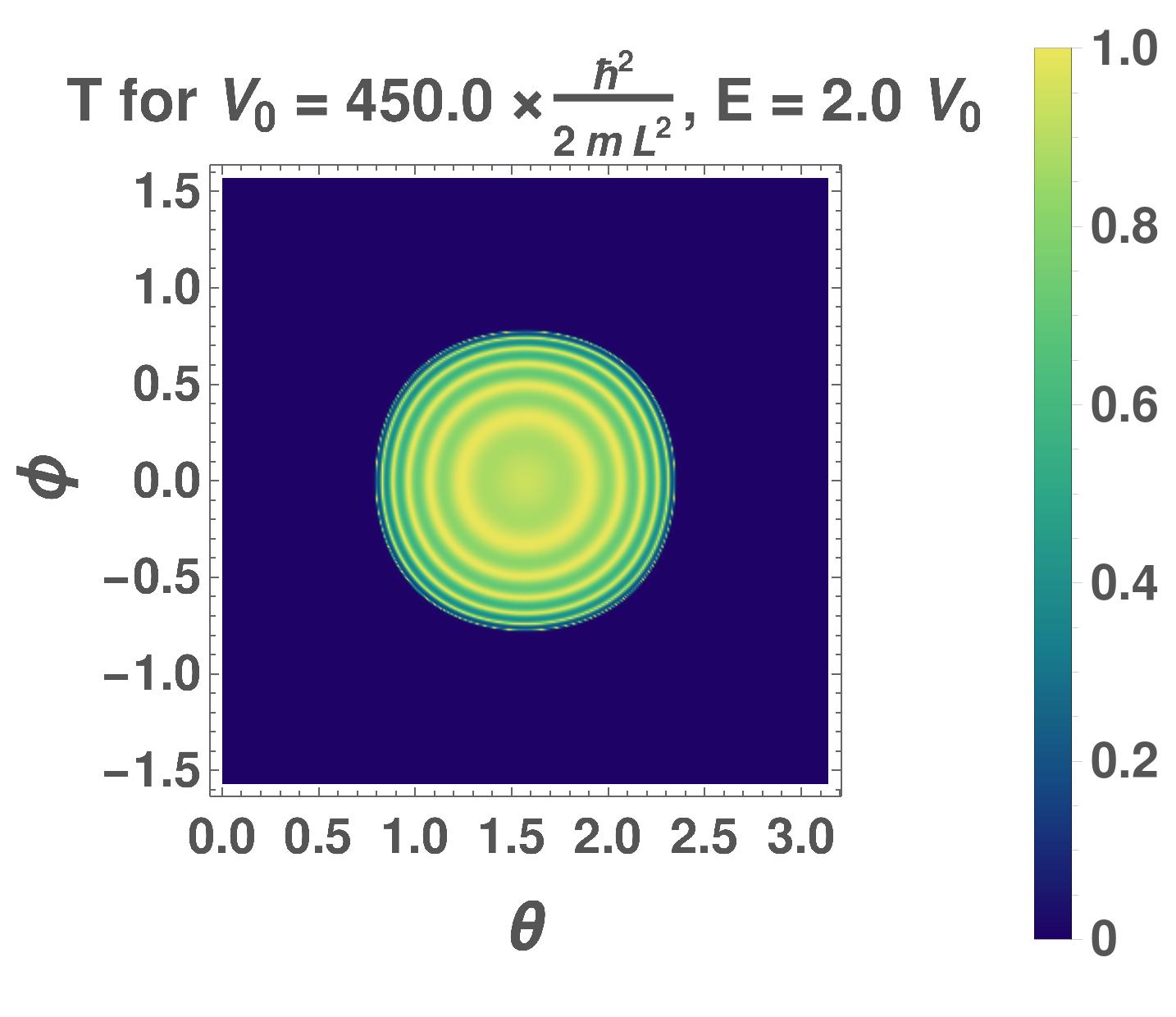}} \quad
\caption{Contourplots of the reflection coefficient ($R$) and transmission coefficient ($T$) for $3d$ QBCP as functions of $(\theta, \phi)$, for various values of $V_0$ and $E$.}
\label{fig3d}
\end{figure}
%%%%%%%%%%%%%%%%%%%%%%%

%%%%%%%%%%%%%%%%%%%%%%%%%%%%%%%%%%%%%%%%%%%%%%%
\subsection{Transmission coefficient, conductivity and Fano factor}

Again, we assume $ W $ to be very large such that $\left (q_{n_y}, q_{n_z} \right)$ can effectively be treated as continuous variables.
Using $ k_\ell = \sqrt{\frac{2mE} {\hbar^2}} \sin \theta \cos \phi,$
$ n_y=\frac{ W\sqrt{ 2 m E }} {h}  \sin \theta \sin \phi ,$ and
$ n_z=\frac{W\sqrt{ 2 m E }} {h}  \cos \theta,$ we get
$ dn_y \, dn_z =
\frac{W^2\times 2\,m \,E } {h^2}  \cos \phi \sin^2 \theta \, d\phi\,d\theta $. Hence,
in the zero-temperature limit, and  for  a  small  applied  voltage, the conductance is given by \cite{blanter-buttiker}:
\begin{align}
G(E,V_0) & = \frac{2\,e^2}{h} \sum_{\mathbf n}   |t_ {\mathbf n,1}|^2 
%%%%%%%
\nonumber \\
 & \rightarrow \frac{ 2\, e^2} {h} \int   |t_ {\mathbf n,1}|^2 \,dn_x\,dn_y
\nonumber \\
& \qquad = \frac{ 4\,\pi\, e^2\,W^2} {\hbar} \left(\frac{ 2\,m\, E } {\hbar^2} \right)
  \int_{ \theta=0 }^{\pi} \int_{\phi=-\frac{\pi}{2}}^{\frac{\pi}{2}}
T( E ,V_0,\theta, \phi) \cos\phi \sin^2 \theta \,d\phi\,d\theta ,
\end{align}
leading to the conductivity expression:
\begin{align}
\sigma (E,V_0)  & = \left( \frac{L }{W} \right)^2 \frac{G(E,V_0)}{e^2/h}
\nonumber \\
&=  8\,\pi^2  \left[ \frac{ E}{\hbar^2/\left( 2\,mL^2 \right)  }\right] 
\int_{ \theta=0 }^{\pi} \int_{\phi=-\frac{\pi}{2}}^{\frac{\pi}{2}}
T( E ,V_0,\theta, \phi) \cos\phi \sin^2 \theta \,d\phi\, d\theta.
\end{align}
Note that there is a twofold degeneracy because we have two
independent conduction band states, and hence an extra factor of two has been included in 
the expressions for $G$ and $\sigma$.

%%%%%%%%%%%%%%%%%%%%%%%%%%%%%%%%%%%%%%%%%%%%%%%%%%
\begin{figure}[]
\subfigure[]{\includegraphics[width = 0.135 \textwidth]{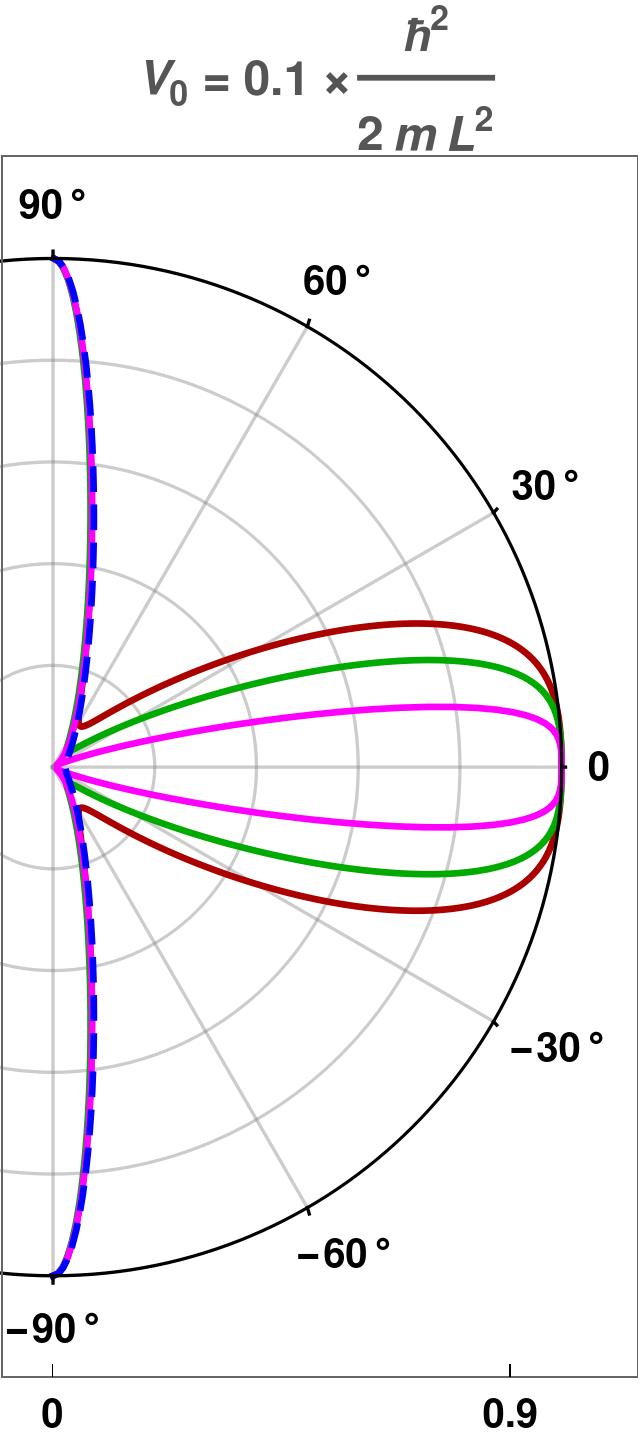}} \hspace{2 cm}
\subfigure[]{\includegraphics[width = 0.14 \textwidth]{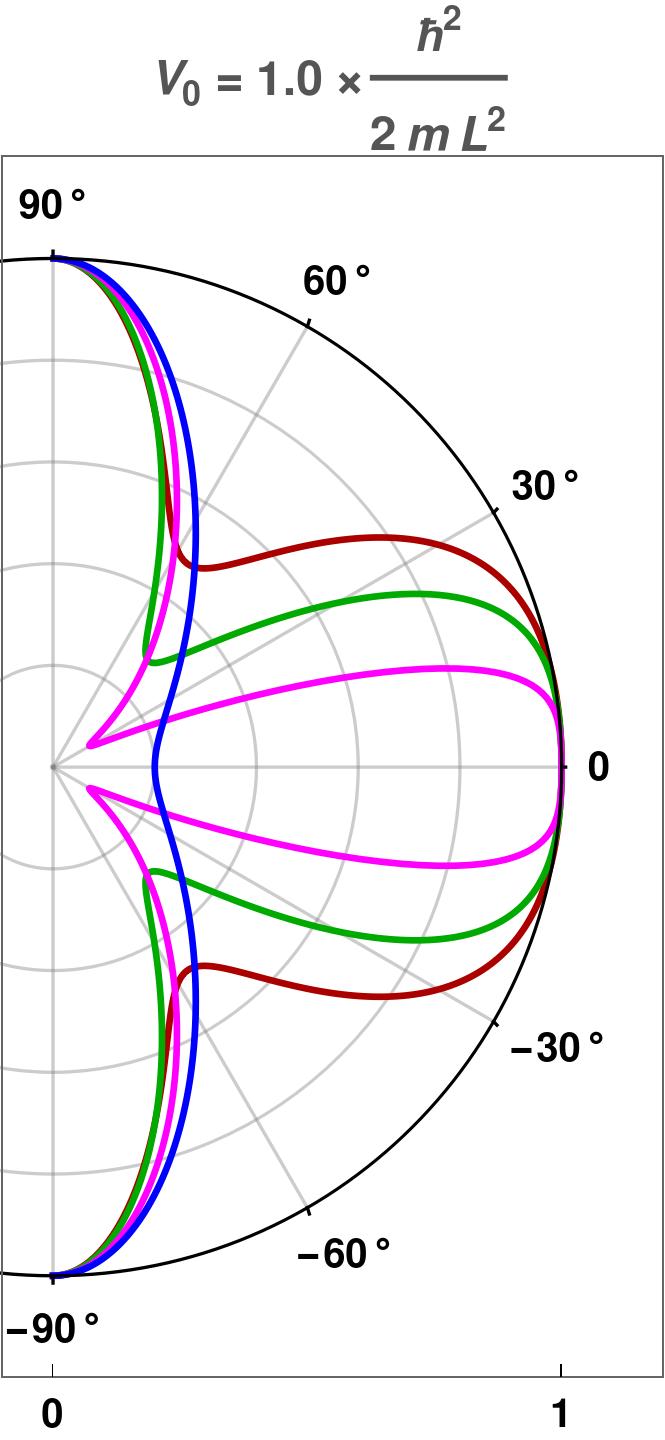}}\hspace{2 cm}
\subfigure[]{\includegraphics[width = 0.14 \textwidth]{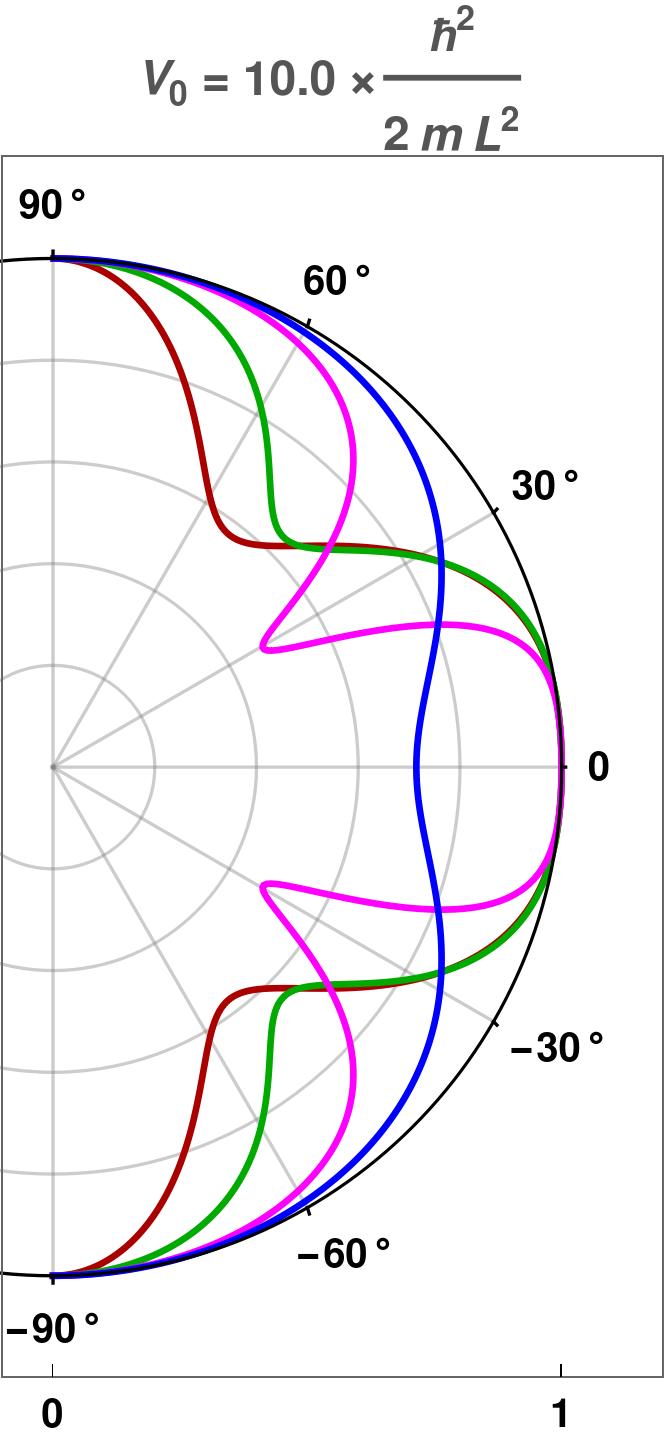}} \\
\subfigure[]{\includegraphics[width = 0.14 \textwidth]{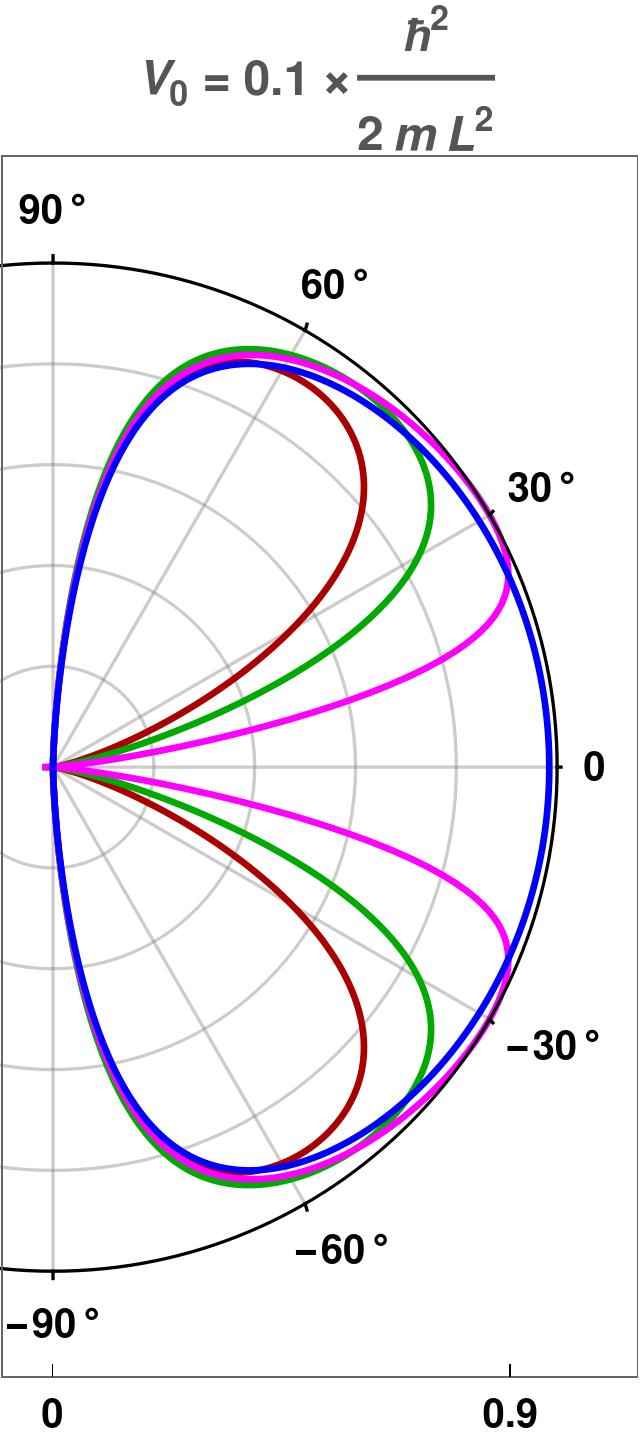}} \hspace{2 cm}
\subfigure[]{\includegraphics[width = 0.14 \textwidth]{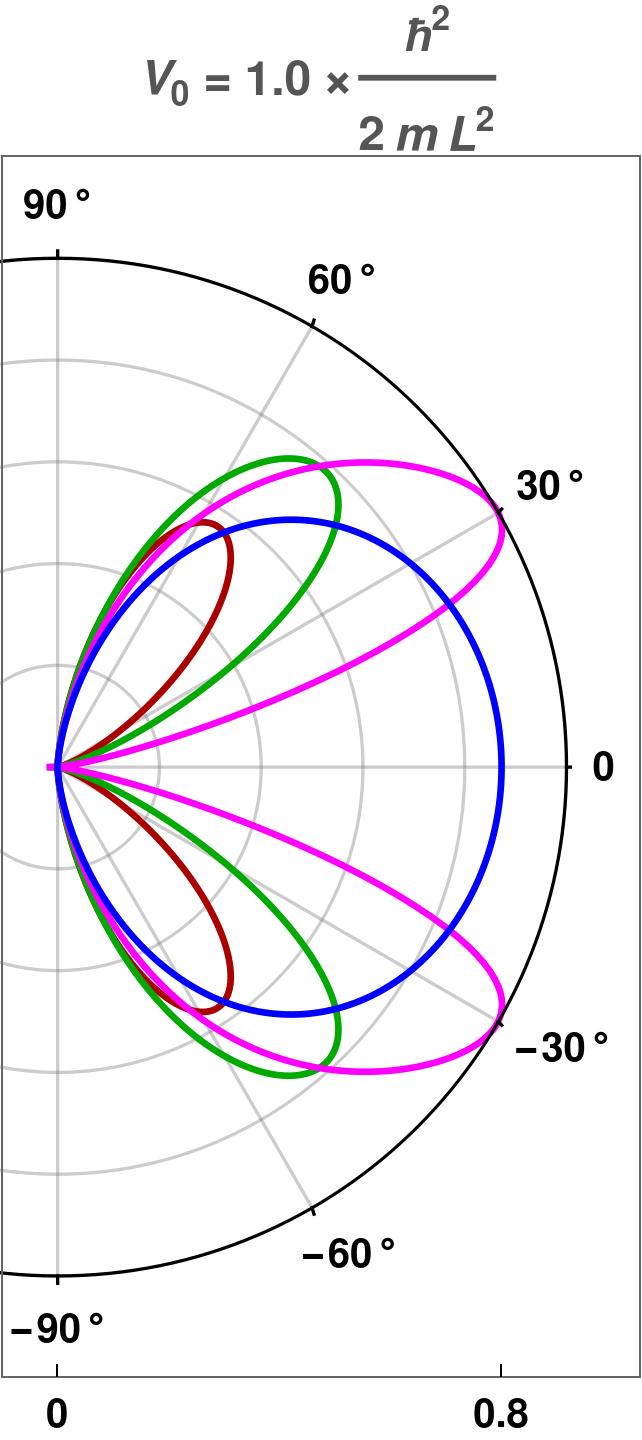}}\hspace{2 cm}
\subfigure[]{\includegraphics[width = 0.14 \textwidth]{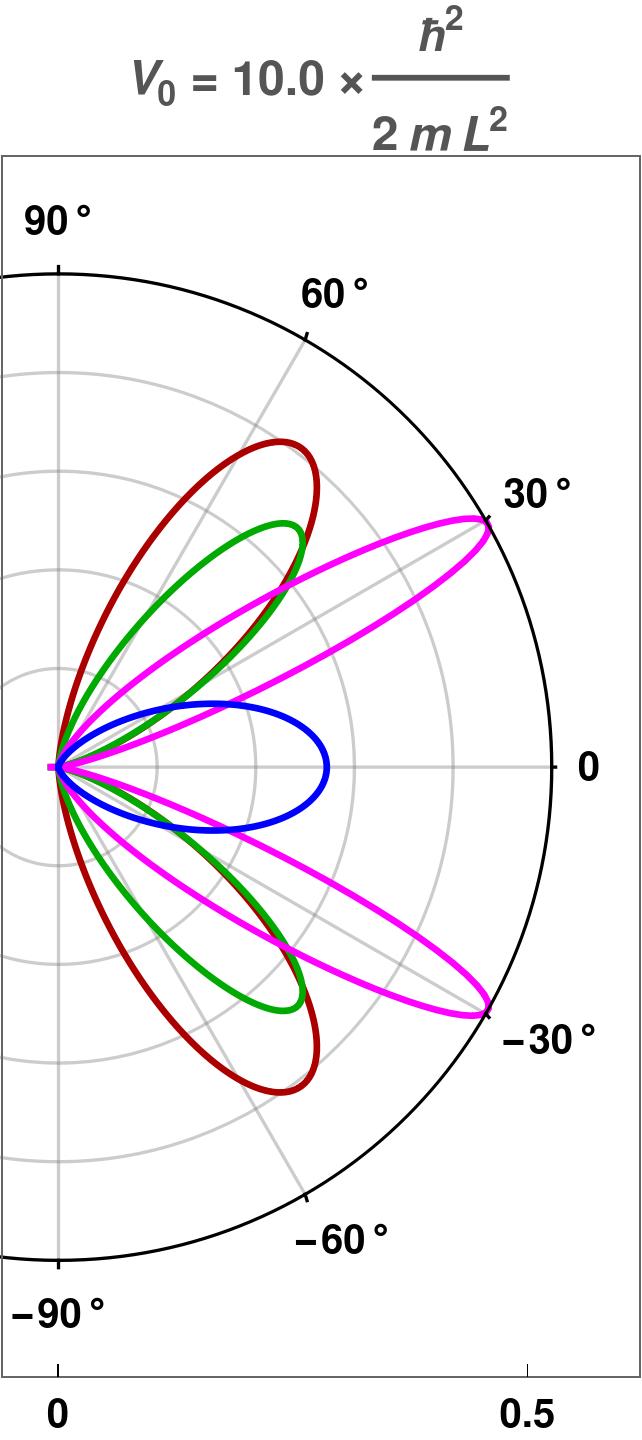}} 
\caption{3d QBCP: The polar plots show the reflection coefficient $ R(E, V_0,\theta,\frac{\pi}{2}) \big \vert_{E \leq V_0}$ and the transmission coefficient $T(E, V_0,\theta,\frac{\pi}{2}) \big \vert_{E \leq V_0}$as functions of the incident angle $\phi$ (in the $xy-$plane with no $k_z-$component) for the parameters $E= 0.3 \,V_0$ (red), $E=0.5\,V_0$ (green), $E=0.8\,V_0$ (magenta)
and $E=1.0 \,V_0$ (blue).}
\label{figRT3d}
\end{figure}
%%%%%%%%%%%%%%%%%%%%%%%%

\begin{align}
 \mathcal{S} (E,V_0) & =\frac{4\,e^2\,\Phi}{h} \sum_{\mathbf n}   
 |t_ {\mathbf n,1}|^2 \left( 1- |t_ {\mathbf n,1}|^2 \right) 
\nonumber \\
&\rightarrow 
\frac{ 8\,\pi\, e^3\,W^2\,\Phi} {\hbar} \left(\frac{ 2\,m\, E } {\hbar^2} \right)
%%%%%%%%%%%%%
\int_{ \theta=0 }^{\pi} \int_{\phi=-\frac{\pi}{2}}^{\frac{\pi}{2}}
T( E ,  V_0,\theta, \phi)  \left[1-T( E , V_0,\theta, \phi) \right] \,d\phi\,d\theta ,
\end{align}
and
\begin{align}
F(E,V_0)  &=\frac  
{\int_{ \theta=0 }^{\pi} \int_{\phi=-\frac{\pi}{2}}^{\frac{\pi}{2}}
T( E ,V_0,\theta, \phi) 
 \left[1-T( E ,  V_0, \theta, \phi) \right]
\cos\phi \sin^2 \theta \,d\phi \,d\theta  } 
{ \int_{ \theta=0 }^{\pi} \int_{\phi=-\frac{\pi}{2}}^{\frac{\pi}{2}}
T( E ,V_0,\theta, \phi) \cos\phi \sin^2 \theta \,d\phi \,d\theta} \,,
\end{align}
respectively. Here, $\Phi $ is the applied voltage.

\begin{figure}[]
\subfigure[]
{\includegraphics[width = 0.45 \textwidth]{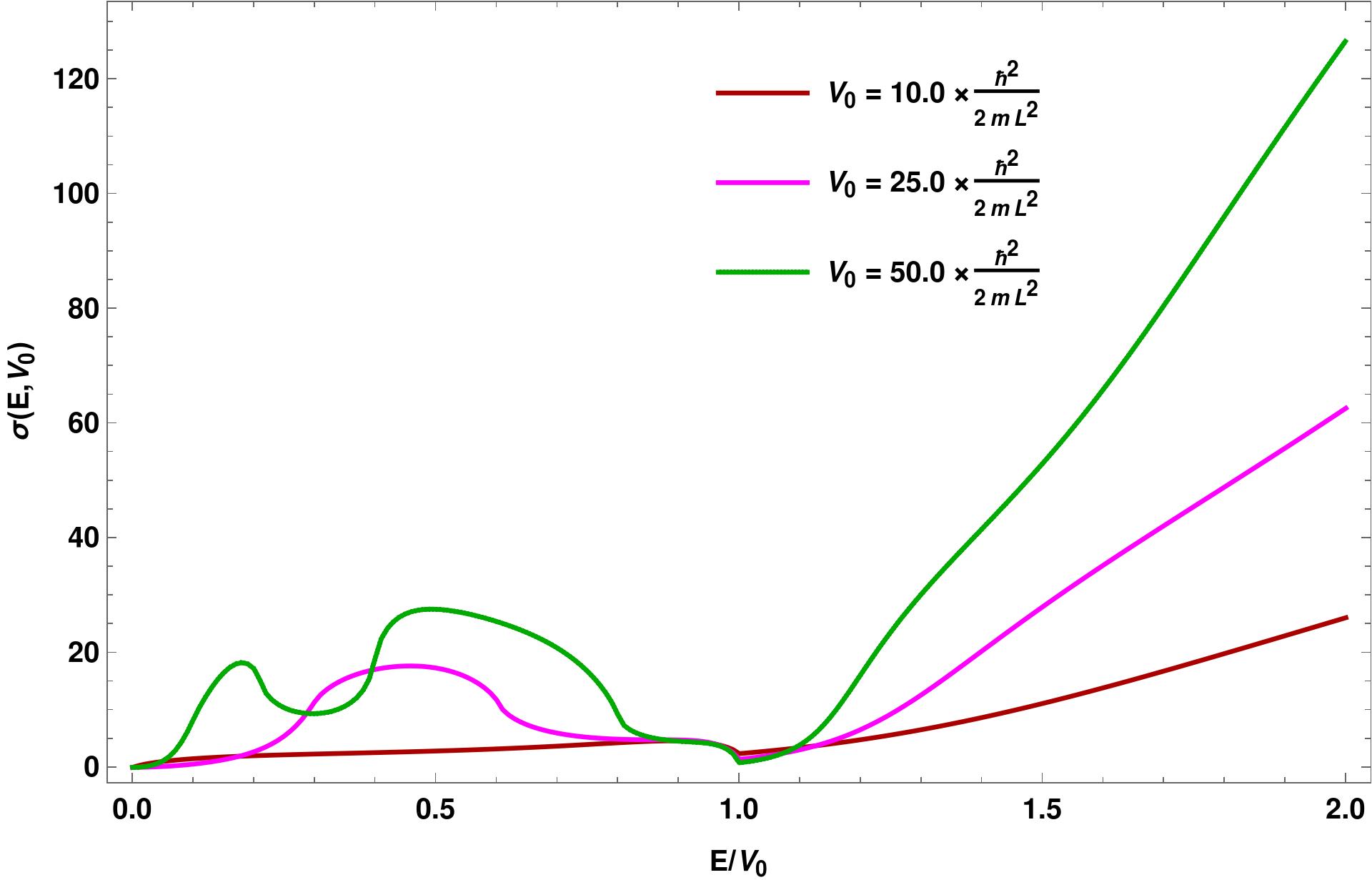}} \quad
\subfigure[]
{\includegraphics[width = 0.45 \textwidth]{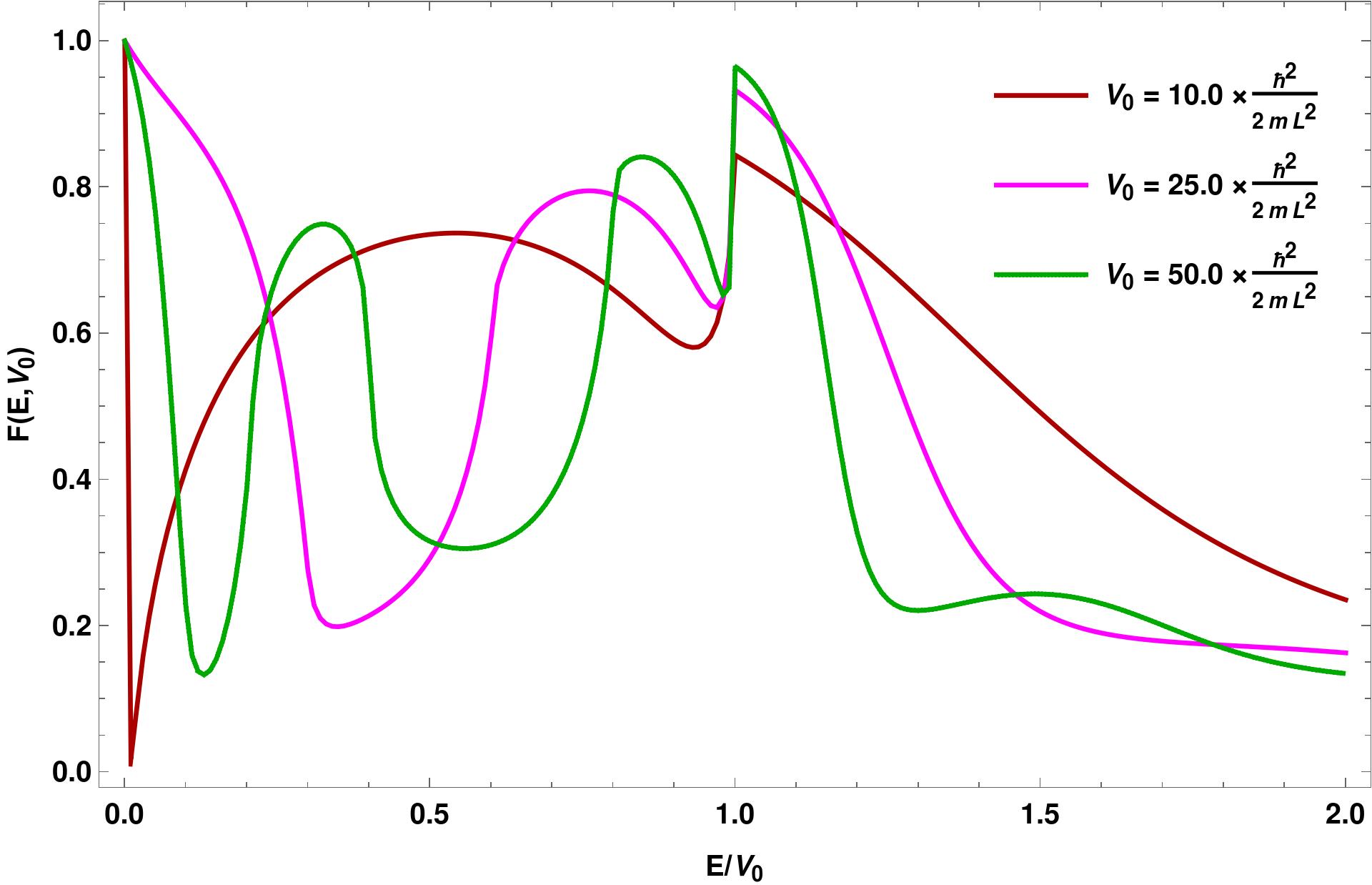}}
\caption{3d QBCP: Plots of the (a) conductivity ($\sigma$ in units of $ 8\pi^2$), and (b) Fano factor ($F$), as functions of $E/V_0$, for various values of $V_0$.}
\label{figfano3d}
\end{figure}

As before, we express $E$ and $V_0$ in units of $\frac{\hbar^2} {2 \,m L^2}$, and study the behaviour of $T( E , V_0,\theta,\phi)$, $\sigma (E,V_0)$ and $ F (E,V_0)$.  
From the expression of transmission coefficient in Eq.~(\ref{eqtval3d}), it is clear that the transmission is zero at normal incidence $(\theta=\pi/2, \phi=0)$, as long as $E<V_0$. This is analogous to the 2d case.
In Fig.~\ref{fig3d}, we show the angular dependence of $ R( E , V_0,\theta,\phi)$ and $T( E , V_0,\theta,\phi)$ via contourplots.
Fig.~\ref{figRT3d} shows the polar plots of $R( E , V_0,\pi/2,\phi)$ and $T( E , V_0,\pi/2,\phi)$ as functions of the incident angle $\phi$ for $E\leq V_0$, which corresponds to $k_z=0$.
Since the transmission coefficient for $E>V_0$ has the same expression both for the 2d and 3d QBCPs, the polar plots of $T( E , V_0,\pi/2,\phi)\big \vert_{E>V_0}$ will be identical to Fig.~\ref{figRTenMoreV}.
In Fig.~\ref{figfano3d}, we illustrate the conductivity $\sigma (E,V_0)$ and the Fano factor $ F (E,V_0)$, as functions of $E/V_0$, for some values of $V_0$.

%%%%%%%%%%%%%%%%%%%%%%%%%%%%%%%%%%%%%%%%%%%%%%%%%%%
%%%%%%%%%%%%%%%%%%%%%%%

%%%%%%%%%%%%%%%%%%%%%%%%%%%%%%%%%%%%%%%%%%%%%%%%%%%
\begin{figure}[htb]
\subfigure[]{\includegraphics[width = 0.14 \textwidth]{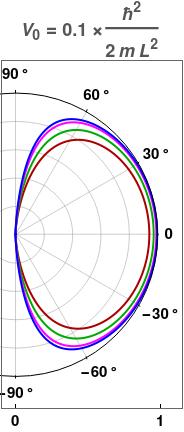}} \hspace{2 cm}
\subfigure[]{\includegraphics[width = 0.142 \textwidth]{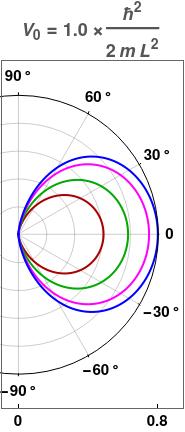}}\hspace{2 cm}
\subfigure[]{\includegraphics[width = 0.14 \textwidth]{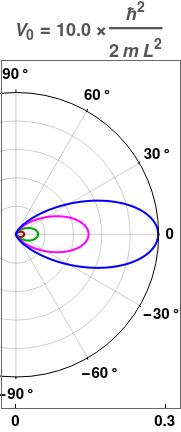}} 
\caption{Normal metal: The polar plots show the transmission coefficient $T(E, V_0,\phi) \big \vert_{E<V_0}$ as functions of the incident angle $\phi$ for the parameters $E=0.3\,V_0$ (red), $E=0.5\,V_0$ (green), $E=0.8\,V_0$ (magenta) and $E= 1.0\,V_0$ (blue).}
\label{figTnormal}
\end{figure}
%%%%%%%%%%%%%%%%%%%%%%%%
%
%%%%%%%%%%%%%%%%%%%%%%%%%%%%%%%%%%%%%%%%%%%%%%%%%%%%
\begin{figure}[]
\subfigure[]
{\includegraphics[width = 0.45 \textwidth]{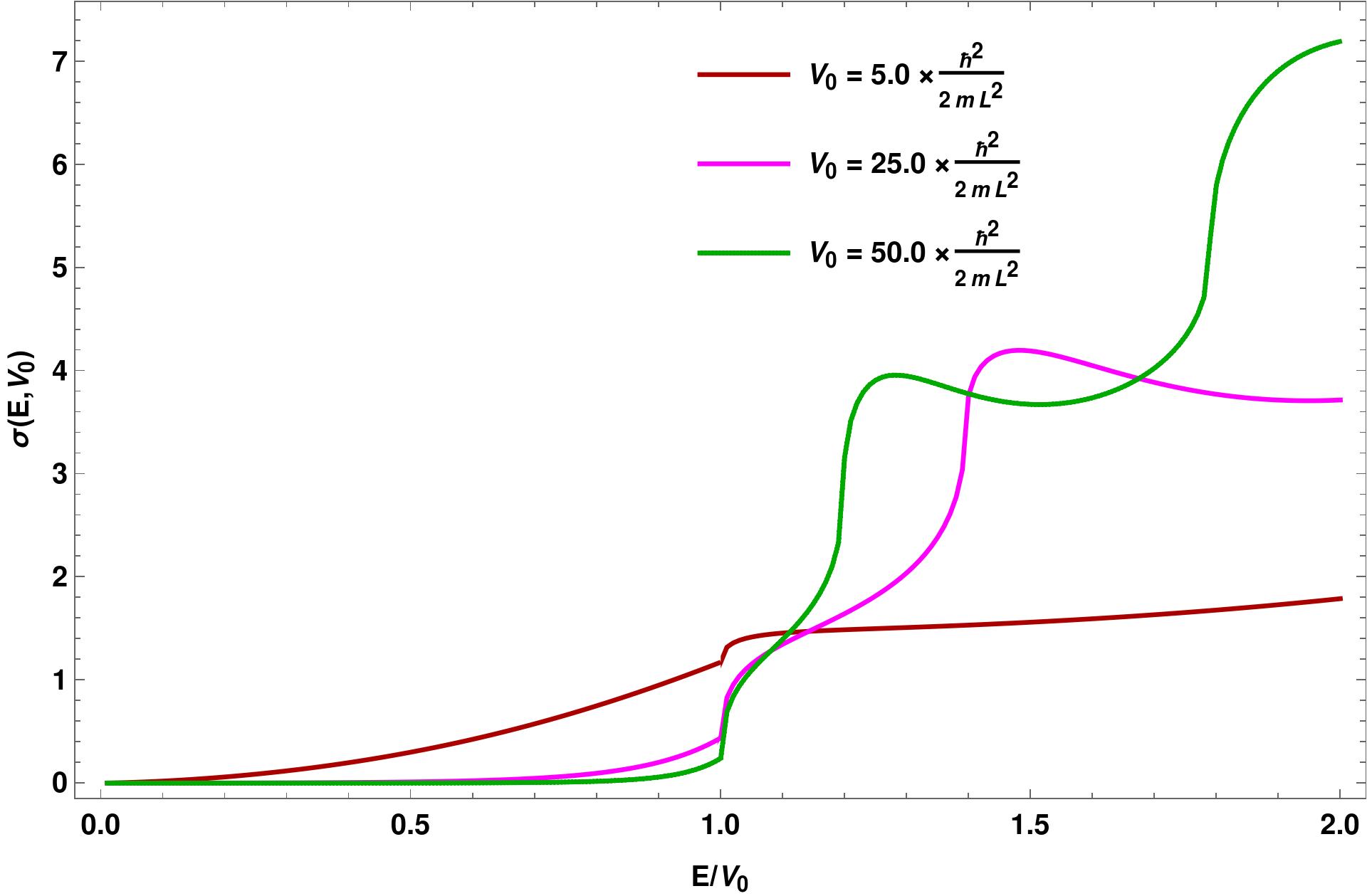}} \quad
\subfigure[]
{\includegraphics[width = 0.46 \textwidth]{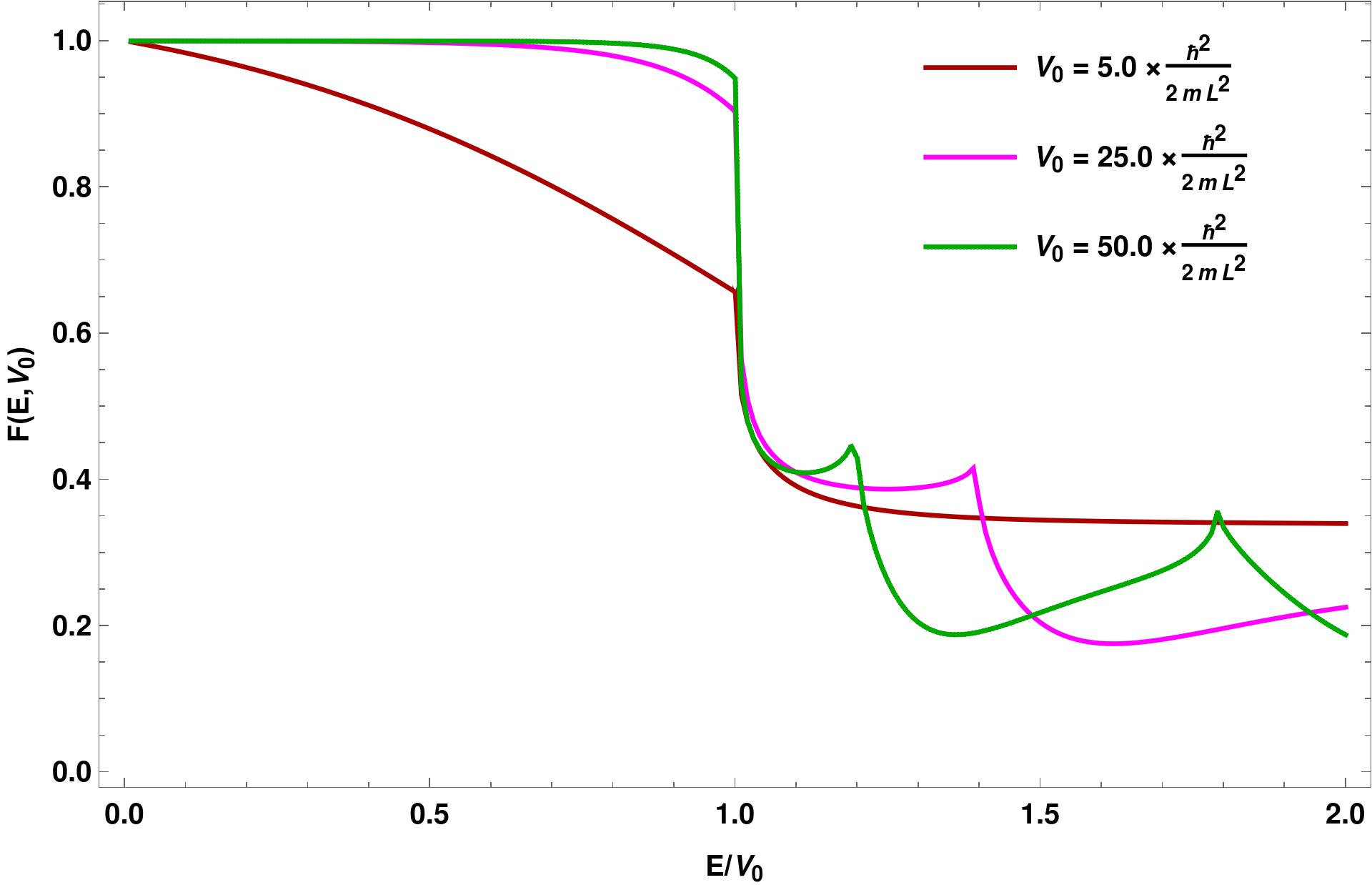}}
\caption{2d normal metal: Plots of the (a) conductivity ($\sigma$ in units of $ 2\pi$), and (b) Fano factor ($F$), as functions of $E/V_0$, for various values of $V_0$.}
\label{figfanonormal}
\end{figure}
%%%%%%%%%%%%%%%%%%%%%%%%
%
%%%%%%%%%%%%%%%%%%%%%%%%%%%%%%%%%%%%%%%%%%%%%%%%%%%%
\begin{figure}[]
\subfigure[]
{\includegraphics[width = 0.45 \textwidth]{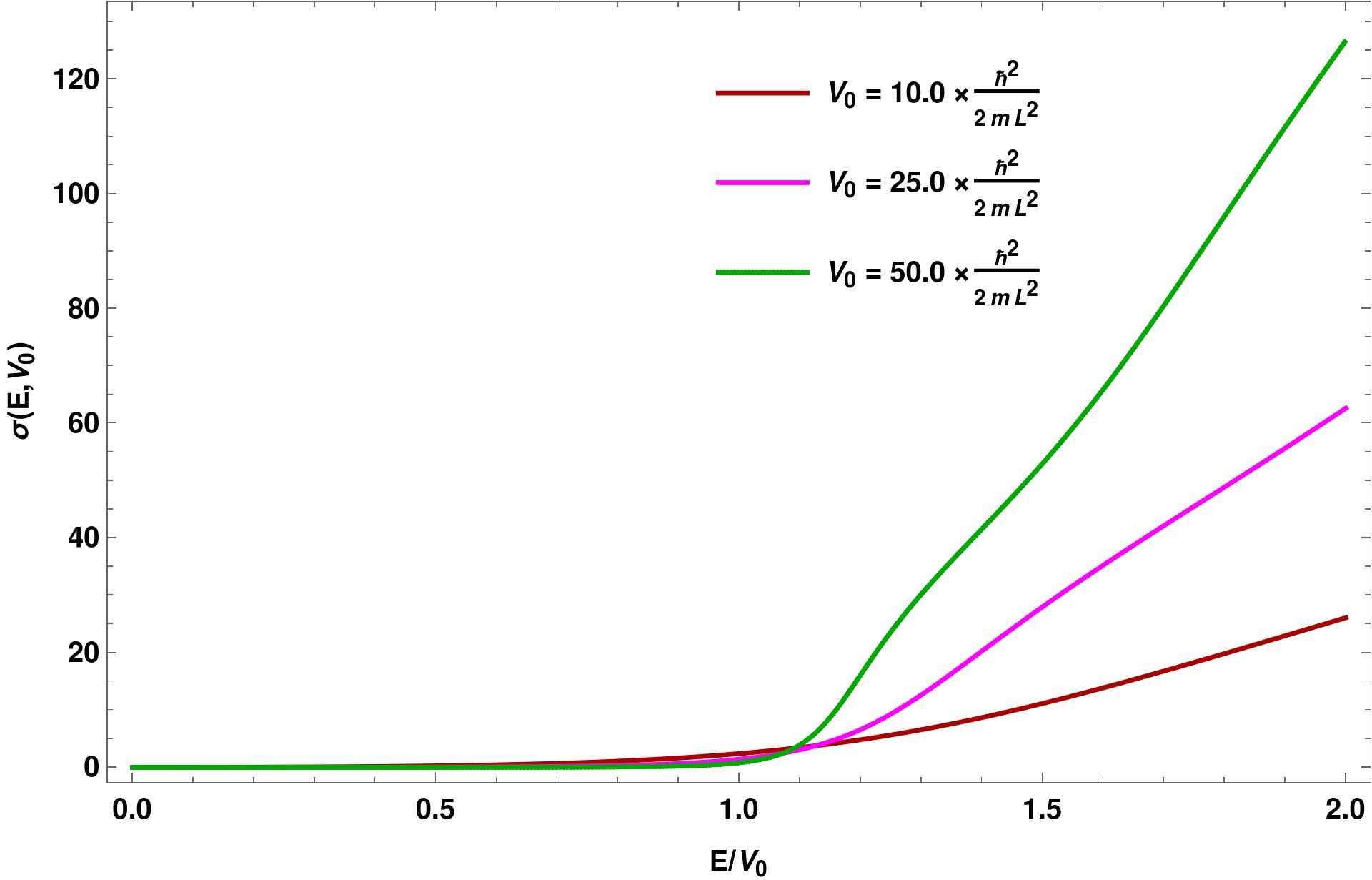}} \quad
\subfigure[]
{\includegraphics[width = 0.45 \textwidth]{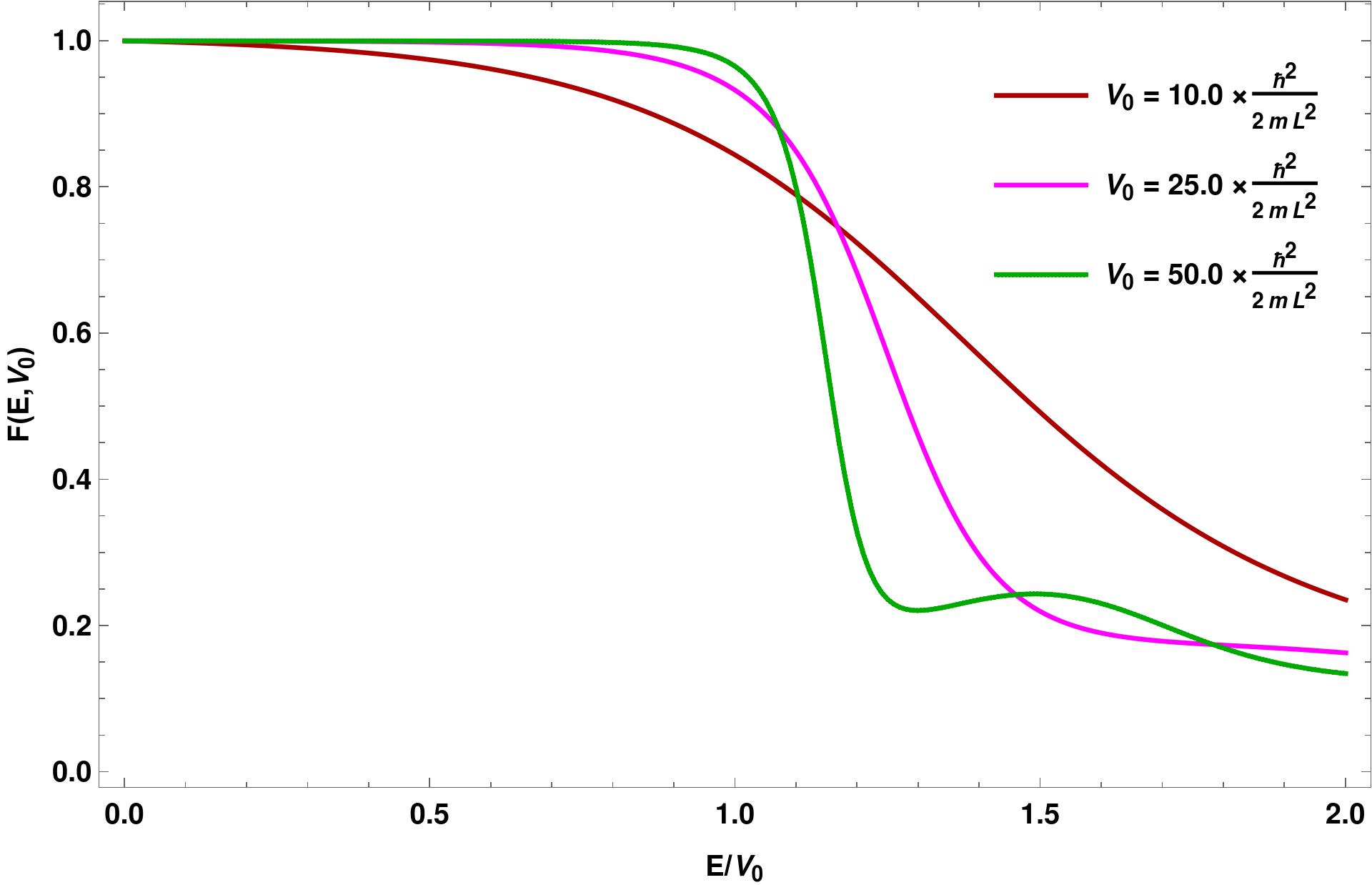}}
\caption{3d normal metal: Plots of the (a) conductivity ($\sigma$ in units of $ 2\pi^2$), and (b) Fano factor ($F$), as functions of $E/V_0$, for various values of $V_0$.}
\label{figfano3dnormal}
\end{figure}
%%%%%%%%%%%%%%%%%%%%%%%

%%%%%%%%%%%%%%%%%%%%%%%%%%%%%%%%%%%%%
\section{Comparison with the the results for electrons in normal metals}
\label{seccompare}

We compare the results obtained for QBCP semimetals with those in normal metals. For normal metals, we have only one electron band to consider where the Fermi energy will intersect (irrespective of the height of the barrier).
Using the continuity of the wavefunctions and their $x$-derivatives at the two ends of the barrier, we can easily find the transmission coefficient to be always given by:
\begin{align}
t(E, V_0) & = 
\frac{2 \,\mathrm{i} \,  \tilde k\, k_\ell }
{\left( \tilde{k}^2+ \, k_\ell^2\right) \sin \left(\tilde{k} L \right )
+2  \,\mathrm{i} \, \tilde{k}\, \, k_\ell \cos \left (\tilde{k} L \right)}\,,
\label{eqtval}
\end{align}
independent of whether $E<V_0$ or $E>V_0$. As expected, this expression varies from the QBCP case only in the $E<V_0$
regime, as whenever $E >V_0$, a quasiparticle excitation moves across the barrier in the same way as a normal metal electron does. 

In Fig.~\ref{figTnormal}, we show the plots of the transmission amplitude $T(E, V_0,\phi) = |t(E, V_0)|^2$ as function of the angle $\phi$, for the normal metal (in the 2d case, or 3d case with $k_z=0$). We also show the behaviour of conductivity and Fano factor for 2d and 3d normal electrons in Figs.~\ref{figfanonormal} and \ref{figfano3dnormal} respectively. As expected, Fig.~\ref{figfanonormal} differs from Fig.~\ref{figfano}, or  Fig.~\ref{figfano3dnormal} differs from Fig.~\ref{figfano3d}, only in the regions where $E<V_0$.

%%%%%%%%%%%%%%%%
\section{Summary and Discussions}
\label{secsum}

From our computations of the tunneling coefficients for the 2d and 3d QBCP semimetals, we have shown that
they exhibit different characteristics than those expected for normal metals. The answers also differ
from those expected for graphene \cite{geim} and three-band pseudospin-$1$ semimetals \cite{fang,zhu}. In particular, QBCPs do not exhibit either Klein or super-Klein tunneling \cite{zhu}. We also note that the transport characteristics for the 2d and 3d QBCP cases show significant differences among themselves. All these observations can be used in experiments to identify the QBCP semimetals.

In future, it will be useful to look at these transport properties in the presence of disorder \cite{rahul-sid,*ipsita-rahul,*ips-qbt-sc} (as has been done in the case of Weyl \cite{emil2} and double-Weyl \cite{emil} nodes) and/or magnetic fields \cite{mansoor,ips3by2}.
Another direction is to examine the effects of anisotropy as well as particle-hole symmetry-breaking terms.
This exercise also needs to be carried out in presence of interactions, which
can destroy the quantization of various physical quantities in the topological phases \cite{kozii,Mandal_2020}.
Yet another direction is to explore the time-dependent transport properties when subjected to a time-dependent potential \cite{zhu-floquet}, using the Floquet scattering theory, and find out if Fano resonance can occur via quasibound states.

%%%%%%%%%%%%%%%%%%%%%%%%%%
\section{Acknowledgments}
We thank Emil J. Bergholtz for useful discussions, and Atri Bhattacharya for help with the figures.

\bibliography{biblio}
\end{document}